\documentclass[preprint]{aastex}

\begin{document}
\baselineskip=11pt

\title{Chandra Observations of Associates of $\eta$ Car:  II. Spectra
  \altaffilmark{1} }
                                                                                     
\author{Nancy Remage Evans, Eric M. Schlegel}

\affil{Smithsonian Astrophysical Observatory, MS 4,                                  
 60 Garden St., Cambridge, MA 02138}

\author{and Wayne L. Waldron }
\affil{L-3 Communications Government Services, Inc., 1801 McCormick Dr., 
Suite 170, Largo, MD 20774}

\author{Frederick D. Seward, Miriam I. Krauss, Joy Nichols, and  Scott J. Wolk }         

\affil{Smithsonian Astrophysical Observatory, MS 4,                                  
 60 Garden St., Cambridge, MA 02138}


\altaffiltext{1}{Based on observations made with the Chandra X-Ray 
Observatory.}


\centerline{Address for Correspondence:}

\centerline{Nancy R. Evans}

\centerline{Smithsonian Astrophysical Observatory}

\centerline{MS 4, 60 Garden St., Cambridge, MA 02138}

\baselineskip=20pt

\begin{abstract}

The low resolution X-ray spectra  around 
$\eta$ Car covering Tr 16 and part of Tr 14 have been 
extracted from a Chandra CCD ACIS image.   
 Various analysis techniques have been applied to the spectra
 based on their count rates. The 
 spectra with the greatest number 
 of counts (HD 93162 = WR 25, HD 93129AB, and 
HD 93250) have been fit with a wind model,
 which uses several components with 
different temperatures and depths in the wind.   
Weaker spectra have been fit with Raymond-Smith models.
 The weakest spectra are simply 
inter-compared with strong spectra.
 In general, fits 
produce reasonable parameters based on knowledge
of the extinction from optical studies and on the 
range of temperatures for high and low mass stars.
Direct comparisons of spectra confirm the consistency
of the fitting results and also hardness ratios
for cases of unusually large extinction in the clusters.
The spectra of the low mass stars are  
harder than the more massive stars. Stars in 
the sequence  evolving from 
the main sequence  (HD 93250) through the system containing
 the O supergiant  (HD 93129AB)
and then  through the Wolf-Rayet stage  (HD93162),
presumably ending in the extreme example 
of $\eta$ Car, share the property 
of being unusually luminous and hard in X-rays. For these
X-ray luminous stars, their high mass and evolutionary status (from the 
very last stages of the main sequence and beyond)
 is the common feature.  Their binary 
 status is mixed, and  magnetic status is still uncertain.

\end{abstract}

\keywords{stars, clusters, X-rays, star formation}

\section{Introduction}

$\eta$ Car and the associated clusters Tr 16 and  Tr 14,
contain some of the youngest and most massive stars in the galaxy.
These stars presumably trace a sequence in descending 
order of mass from the  enigmatic 
$\eta$ Car, the Wolf-Rayet star HD 93162, the O3If 
(presumed Wolf-Rayet precursor) HD 93129AB,  HD 93250 (O3 V), 
and on through other main sequence stars.

An early Chandra image was centered on $\eta$ Car, and 
contained Tr 16 and part of Tr 14.  $\eta$ Car itself from this
image has been discussed by Seward et al. (2001).  The
luminosities and hardness ratios of  other stars
in the clusters have been discussed by Evans et al. (2003, hereafter 
Paper I).  Source variability will be discussed in a future (in 
preparation.)

 The Chandra ACIS image provides low resolution spectra
 of sufficiently luminous  sources.
 This paper is a discussion of 
the spectral properties of these stars.
  The low resolution spectra
supplement grating spectra which can only be obtained for the 
very brightest X-ray sources. 
 HETG spectra have been obtained for $\eta$ Car itself
(e.g. Corcoran et al, 2001). 

A major aim of X-ray studies of this region is to understand the factors that
dominate the evolution of the most massive stars in the galaxy. The large
number of O stars in this part of the sky justifies the use of less than
optimal Chandra data (see section 2). Since the surprising discovery of X-rays
in O stars with the Einstein satellite 
(see Paper I for background), it has been known
that at least in some cases, X-ray emission originates in shocks in the strong
winds produced by very massive stars. As shown by MacFarlane et al. (1991), a
stellar wind distribution of shocks is expected to produce a unique X-ray
emission line signature of very broad, highly asymmetric, blue-
shifted X-ray emission line profiles. Although Chandra HETGS observations do
show the expected broadness, the emission lines are found to be essentially
symmetric and unshifted (Waldron and Cassinelli 2001; Miller et al. 2002). The
one exception is the early O-star $\zeta$ Pup (Cassinelli et al. 2001) which
shows blue-shifted lines, but the predicted large line asymmetry is 
minimal.  The picture may be further complicated by the
possibility that in some cases a binary system may contain two massive stars
giving rise to "colliding wind" systems. Recent investigations (eg Schulz et
al. 2003) add a further complication in that some hot stars appear to produce
X-rays through magnetic processes (eg. $\theta^1$ Ori C, A, and E). 
 Tr 14 and Tr 16 therefore provide one of the best locations to
investigate X-ray production in a broad range of massive stars.

General introduction to the clusters 
Tr 14 and Tr 16 is provided in Paper I.  
Since that paper was submitted, two additional 
studies on the area have become available 
which are particularly relevant to our discussion.  Raassen et al
(2003) have analyzed an XMM-Newton high-resolution spectrum of 
 WR 25.  Albacete et al (2003) have discussed the luminosities 
and low-resolution spectra of other sources 
in the region, also using XMM-Newton data.  We will include these 
papers in our discussion later in this paper.

X-rays are particularly good tracers of activity high in the atmospheres of
stars. In high mass young stars, the mass loss from the atmosphere is
fundamental to the future evolution of these objects. It is also linked to the
dynamical and chemical history of the interstellar medium. In this paper
(Paper II), we focus on the information that ACIS-CCD low-resolution spectra
provide. We are investigating three areas in particular. First, what insights
can the spectra provide to the factors effecting evolution of the most massive
sequence of objects ($\eta$ Car, HD 93129AB, WR 25, and HD 93250)? Second, how
are the X-ray spectra and intrinsic photospheric stellar parameters (such as
temperature) or extrinsic parameters (such as reddening) related for the O
main sequence stars from O3 through O8? Finally, what are the characteristics
of the X-ray spectra for the cool (less massive), presumably pre-main sequence
stars? ``Cool stars" refers to the photospheric temperature of low mass
stars, even though their X-ray temperatures are often higher than those of
high mass stars. Specifically, our analysis consists of fitting the spectra
to multi-component models (where the spectra have enough counts), fitting
simpler CIAO/Sherpa models to weaker spectra, intercomparing the spectra
themselves in appropriate subgroups, and comparing derived temperatures and
hardness ratios. 

 Succeeding sections of this paper discuss the following
topics. 1. The sources are identified. 2. Strong sources are modeled with a
detailed wind model. 3. Weaker sources are fit with models available in the
CIAO SHERPA package. 4. Spectra of sources are intercompared. 5. Information
which can be obtained from ACIS low resolution spectra is discussed. 6.
Results are summarized. 

 \section{Spectra} $\eta$ Car was imaged using the ACIS-I detector 
 onboard Chandra (Garmire et al. 2003).  Two observations were obtained 
 (OBSID's 50 and 1249) within the space of 1.2 days.  There data
 were merged by reprojecting all the events into a single 
 reference frame.  Spectra were extracted for
sources with a reasonable number of counts, more than 100 if possible.
Background subtraction was handled by extracting counts in an annulus
surrounding each source to account for local variations in the diffuse X-ray
emission. Any contamination from point sources intruding on the background
annulus was eliminated before extracting the background. In a few cases, we
went below our 100-count limit. For example, weaker sources 
were used  for cool stars
to obtain representative examples of the class. The stars
with extracted spectra are indicated on our Chandra image shown in Fig 1. To
avoid confusion, sources M and N have not been marked. They are the two
sources immediately south of source H. The sources are listed in Table 1
(ordered by decreasing optical brightness), which lists the alphabetical
identification in Fig 1. Successive columns are the alphabetical
identification, the star name, the source number from Paper 1, RA (2000),
Dec(2000), the counts, two hardness ratios defined below (HR$_{MS}$, the soft
hardness ratio and its error, and HR$_{HM}$, the hard hardness ratio and its
error), the V magnitude, the spectral type and the E(B-V). These quantities
are taken from Paper I, with the final 3 columns generally adopted from Massey
and Johnson (1993, referred to as MJ93 below). Hardness ratios are formed from
counts in bands 0.5 to 0.9 keV, 0.9 to 1.5 keV, and 1.5 to 2.04 keV. HD
93129AB is a particularly complicated source (unresolved in our image). The A
and B components have spectral types O2If* and O3.5 V((f+)) respectively
(Walborn, et al. 2002). In addition, component A was recently resolved by
Nelan and collaborators 
(Paper I; Benaglia and Koribalski 2004), resulting in an
additional component 0.9 mag fainter than A.

The ACIS detector suffered radiation damage immediately after the
opening of the outer doors that protected the instrument during
launch.  The radiation damage increased the charge transfer
inefficiency (CTI) of the ACIS CCDs with the effect of creating a gain
shift to lower energies as well as a reduction of the spectral
resolution.  These effects are a function of the distance from the
original location of the charge (event) to the CCD readout node.  The effect
is non-linear (Townsley, et al. 2000).

  The observation of ${\eta}$ Car occurred near the middle of the maximal 
change in CTI.  Furthermore, the operating temperature of ACIS during this
time was -100$^{\circ}$ C, a temperature for which minimal calibration 
data exist.  As a result, spectral fits may be uncertain which we explore
in some detail.  We judge the possible CTI-induced differences between the
two OBSID's to be smaller than the uncertainties in the calibration.  We 
reach this judgment solely because the observations were contiguous 
in time.

  We approximated an instrument response matrix by interpolating between
a matrix at -90$^{\circ}$ C and one at -110$^{\circ}$ C, and introduced
a correction for CTI as described below\footnote{Current processing 
software correct for the CTI for a focal plane temperature of -120$^{\circ}$,
however, the data we  used in this analysis do not include these corrections 
because they were not obtained at that temperature.}.

We can quantify the impact of CTI using three bright stars in the $\eta$ Car
field (sources D, E, and F).  We extracted the spectra of these
sources and found  one consistent feature across the spectra: an
apparent absorption line near 2.1 keV.  The line is an absorption
feature in the iridium coating on the mirror, so it is intrinsic to
the instrument and is impressed upon all the spectra.  The limitation
in this approach is the presence of only one bright source (D);
sources E and F are weaker.

We modeled the absorption as a gaussian on a sloping background.  The
fits yielded the line center and line width and are shown in
Fig. 2.
 The top sub-plot shows the three fits overlaid; it is
clear that a gain shift progression occurs from source F to source E,
with the minimum of the absorption moving to lower energies.  The
gaussian width becomes larger in the same direction.


  Fig. 3 summarizes the fit parameters, now as a function of the number
of pixels from the mean position of the source to the readout node 
measured along a readout column.  The local gain shift and decreased
spectral resolution resulting from CTI are apparent.  Local effects of
CTI may be corrected by the use of individual response matrices.

  All spectral models included a global gain shift with a linear slope of
0.95 and an offset of 0.0.  These values were taken from the analysis of
Seward et al. on ${\eta}$ Car itself also using the 2.1 keV iridium feature.
${\eta}$ Car is a much brighter source with better count statistics, so 
the global shift determination, obtained by requiring the Iridium feature 
to lie at its laboratory energy, is accurate.  These values were used
during the spectral fits to the sources described here and fixed at the
values listed above.   We estimate that the combination of global plus
local gain shifts reduces the error on the gain to about 10\% or less.

 Response matrices and effective area files were
then generated for each source in Table 1 individually.  For the generally low
count rates of these OB stars, this approach to calibration provides
spectral parameters we estimate are accurate to better than 25\%.  The
largest discrepancies will occur at the lower energies; the high
column toward the OB stars near $\eta$ Car largely eliminate the soft
events, however.

To confirm this approach, we have compared the fits to our 
spectra with fits to the same sources
in an XMM observation (Albacete Colombo, et al., 2003)
in Table 6. As discussed in Section 6.1, we find the fits differ in 
some cases because XMM constrains the extinction better because 
of its softer response.  When the fits produce the same column 
density, the luminosities are also similar.  Unfortunately, we 
are unable to use zero order data from several HETG spectra to 
confirm our analysis.  Only one of the strong sources falls 
on the BI ACIS-S CCD chip, and the relatively few counts would not
be a good test of our correction.

\section{Strong Sources: Wind Absorption Models}

     Chandra HETGS observations have finally verified that the X-ray emission 
from O stars
arise from various regions distributed throughout their stellar winds 
(e.g., Waldron and Cassinelli
2001) which implies that all X-ray spectral model fitting efforts must
incorporate the effects of
stellar wind absorption.  The first attempts to explore wind absorption
effects were carried out
during the Einstein era (e.g., Cassinelli and Olson 1979; Waldron 1984).
 In these studies it was
shown that although the wind absorption cross sections are significantly
smaller than the cold ISM
absorption cross sections below 1 keV, since the wind column densities
can be significantly larger
than the observed ISM column densities, the overall observed X-ray
spectral distribution is
predominately controlled by the absorption properties of these stellar
winds. Even when the ISM
column density is comparable to the wind column density, the effects
of wind absorption still
play an important role in the interpretation of observed X-ray spectra
(e.g., Waldron et al. 1998).

The stellar effective temperature (strength of the UV radiation field) is the
dominant factor that determines the energy distribution of the wind absorption
cross sections. These cross sections are also affected by the density
distribution of the wind but these effects are most noticeable only at
energies below $\sim$ 0.1 keV. In the ACIS-I energy band ($>$ 0.1 keV) the
effects of density are minimal. For our stars with the greatest number of
counts (HD93129AB, HD93250, and HD93162), effective temperature range between
$\sim$ 50000 to 54000 K (Howarth and Prinja 1989; Lamers and Leitherer 1993),
so their wind cross sections are very similar. Figure
4 shows a comparison of the wind cross sections with the ISM cross sections
for these stars. Since the wind is subjected to a strong UV radiation field,
the dominant H and He absorption cross sections seen in ISM cross sections are
greatly reduced revealing a highly structured collection of wind cross
sections below 1 keV from lower abundant elements (e.g., CNO). For example,
the energy shift of the O K-shell edge illustrates that the wind is in a
higher state of ionization which leads to the commonly used phrase "warm-wind"
absorber model. A slight difference exists between in our Fig 4 and
Waldron et al. (Fig 2) because Fig 4 is based on an effective temperature of 
$\sim$ 52,000 whereas, the Waldron et al 
plot is based on an effective temperature of 40,000.

     Following the wind absorption X-ray fitting procedure outlined by 
     Waldron et al. (1998),
we obtain wind absorption model fits to our three ACIS-I spectra using 
the MEKAL emissivity
model (of Mewe, Kaastra, and Liedahl). 
The ISM column densities are fixed parameters in our model fits.  The 
adopted values are given in Table 2 and were determined from their E(B-V) 
values given in Table 1 as described in Shull and Van Steenberg (1985).             
 (The fits in Sections 3 and 4 are made for spectra merged between
two 10 ksec images.)
We start by assuming a single component model and add components 
where necessary to
obtain statistically better fits.  All single component fits were clearly
 not able to explain the
spectral distributions.  One star (HD93250 = Source E) only required two components, 
whereas, the other
two (HD 93129AB = Source F and HD 93162 = Source D) 
required three components.  The fitting parameters 
are given in
Table 2.  These fits include the CTI correction as determined in  
 Sec. 2.  To handle this calibration issue 
we developed a special
algorithm to apply artificial broadening and energy shifts to all lines.  
The algorithm extracts the
line energy and strength from the input emissivity.  Each line is then 
characterized by a Gaussian
line profile with the appropriate line broadening  (HWHM in Table 2) 
and energy shift. 
This modified emissivity is then folded through the ACIS-I RMF and ARF to 
obtain our best fit
models.  We point out that this initial approximate approach to address 
the calibration problems
assumes that all lines are subjected to the same broadening and shifts.  
In addition, the continuum
and cross section energies are not shifted, which does not appear to be an 
important effect. 
Although the broadening and shifts do tend to improve the  $\chi^2$, from a 
statistical point of view
the difference is essentially minimal.  However, a visual inspection of 
the resultant best-fit spectra
clearly show that broadening and shifts indicate better fits to spectra.

The following provides additional information about the parameters 
in Table 2.  
     For the variables T$_X$, N$_W$, and EM, the quoted best-fit values 
     and errors are based on
taking the  $\chi^2$ 90\% confidence range for a given variable and 
averaging to obtain the best-fit
value.  The error is then associated with the difference.  All variable 
upper or lower limits
represent the upper or lower limits on the 90\% confidence range.
     The errors on the observed flux, F$_O$, and intrinsic flux, F, for each component are
determined by comparing the model predicted counts with the observed 
counts in specified energy
bands.  For each star these energy bands are dependent on the location
at which each component is expected to
provide maximum flux.  The relative errors in observed and predicted 
counts are then compared to
the corresponding observed count statistic error, and
the adopted error is the larger of the two.

     The results of our fits, shown in Figs. 5--7,
 indicate that all three targets show evidence 
     for a hard X-ray
component with a temperature $\ge$ 10 MK which appears to be deeply embedded 
within the stellar
wind  ( with a column density N$_W$ $>$ 10$^{22}$ cm$^{-2}$).  
The softer components, based on the derived N$_W$, 
all appear to be
distributed farther out in the wind.  Although this distribution appears 
consistent with a shock
model, the derived temperature distribution (increasing inward) may be a 
problem for a shock
model scenario as discussed by Waldron and Cassinelli (2001).

\section{Weak Sources}

The remaining sources fall into two groups: a group for which spectral
fits are potentially meaningful (sources B = -59$^o$2600, 
C = HD 93205, and G = HDE 303308) and a group for
which spectral fits yield little more than a temperature {\it or} a
column density (sources I, J, K, M = -59$^o$2635, and N = HD 93343).

We fit the spectra of             
the first group with three types of Raymond-Smith models: all                   
parameters free to vary, freezing the plasma energies of the O stars at         
0.4 keV, and freezing the column densities at $0.3 \times                       
10^{22}$~cm$^{-2}$. The optical
extinction typical of the hot stars corresponds
to this column density and represents a minimum extinction where
there is no additional circumstellar wind absorption. The adopted 
temperature is confirmed by the fit in Table 4 to source C, the strongest
of the normal O stars.  
  The fits were carried out using the fitting codes {\tt
Sherpa} and {\tt xspec}. 
Figure 8
shows the results while Table 3
lists the model parameters. In Table 3 we include the formal errors 
from the fits, although intercomparison of the results shows 
that the actual uncertainties are larger than the formal errors.
 We have suppressed the error bars in the
upper portion of Fig. 8 for legibility.  The lower plot shows the
residuals of the fit to the spectrum of source C; the residuals for
sources B and G are similar in vertical range, but contain fewer data
points spanning the energy scale.  Sources B and G
(O6 V and O3 V respectively) have nearly
identical spectral shapes (and parameters) and differ from Source C
(O3 V) in
that C shows evidence for emission lines (Table 4).  The
lines in C were all fit using zero-width gaussians and are significant
at $>$99.7\% for 2 parameters of interest (line center, line flux).
Two of the features fit with Gaussians (0.579 and 0.892 keV)
match  strong features seen in grating spectra of hot stars
(e.g. $\zeta$ Pup, Cassinelli, et al., 2001);
one of the gaussians (0.686 keV) did not have a match.  Thus, the 
low resolution spectra pick out some emission lines accurately, 
but not all.  The temperature and absorption for source C in 
Table 3 are not exactly the same at that in Table 4. However,
when either of the parameters is frozen at  a value similar to 
those in Table 4, a similar combination is produced.  This is a  
good illustration of the non-uniqueness of the fitting procedure
(even where the $\chi^2$ can be used to select between fits).
In Table 5 (below) two temperature fits were made.  For Source C, 
the two temperatures bracket the single temperature in Tables 3 and 4.  

For the weakest sources, the spectral fits essentially permit only an
adjustment of the model normalization and perhaps one additional parameter.
Within the errors, it is largely irrelevant whether one fixes the
model temperature and fits the column or vice versa.

As part of our exploration of the information that can be obtained from 
ACIS resolution spectra, we have also fit Raymond-Smith models to the 
strongest spectra, those fit with the wind-models (Table 5). Experience
has shown that  the absorption needs to constrained in the fits.  
 Because the 
absorption is reasonably constant throughout the clusters, we have 
fixed the column density at  $\sim$0.3$\times$10$^{22}$                
cm$^{-2}$ as in Table 3. The simplest model combined with this extinction
that provides reasonable fits to the spectra is a two temperature 
model.  One temperature fits result in poor reduced chi-squared 
values, as seen in Table 3.  (The errors in Table 3 are poisson errors
from the solution plus 25\% because of the calibration uncertainty
added in quadrature.)
 Table 5 provides fits for the  two temperature models
for sources C, D, E, and F.   The reduced $\chi^2$ is markedly 
smaller in the two temperature fit than for a single
temperature.  These temperatures 
will be discussed in conjunction with the physical parameters of the stars.  

    When we compare the temperatures in Table 3 and those in Table 5, 
several features emerge.  Table 3 is partly an exploration of the 
robustness of the fitting results, to see which values are returned 
most consistently when  different parameters are held constant.
The fit which fixes the absorption in Table 3 is the most similar 
to the fit in Table 5. For these fits, the single temperature in Table 3 always
falls between the two temperatures in Table 5.  Furthermore 
the  mean temperature in Table 5 and the absorption constrained fit
in Table 3 order in the same way, C, E, F, and D from  the coolest 
to the hottest.  For the two stars  most likely to have 
significantly more absorption than the interstellar absorption, 
source D = WR 25 and source F = HD 93129AB, the reduced $\chi^2$  in
Table 3 shows 
a marked increase when the absorption is held to the ISM absorption.

\section{Comparisons}

 An exploration of the spectral information independent of response 
matrix or calibration issues is justified because of 
 the large number of O stars in the region.
 Tr 14 and Tr 16
contain many massive stars (including several
O3 stars, the hottest spectral types).  In addition, ground-based 
observations have provided extinction measures for these stars.  
In this section, we provide direct comparisons between spectra with a 
variety of optical parameters.  

\subsection{O3 Stars} Fig. 9 shows the comparison of the spectra of 3 O3 V stars
in the image. Source E = HD 93250 is the most X-ray luminous normal star in the
Chandra image and we use
it as the reference spectrum in the comparisons. In Fig 9,
in order to compare spectral slopes, sources C and G
have been scaled so that they approximately match source E between 0.8 and 1
keV. The figure shows that the spectral slope from 0.8 to 2 keV is the same for
sources C and G, however, source E is clearly harder.
 Sources C and G have relatively more counts
between 0.5 and 0.8 keV than source E. Thus the brightest source in X-rays also
has the hardest spectrum. From the optical results, all 3 stars have similar
E(B--V) (Table 1). Thus the fact that source E has the
hardest X-ray spectrum should not be caused by differential absorption, but
apparently reflects a different source temperature. It is interesting to 
note that source E with the hardest spectrum is not known to be a binary
(Paper I), but the softer source C is a binary, and source G may be a 
binary. 

 \subsection{Absorbed O3 Star} Fig. 10 shows the comparison of the standard O3
 X-ray spectrum (Source E; solid line) with the O3 star (Source L; squares)
 which has a high optical absorption [E(B--V) = 0.94 mag, Table 1]. Source L
 lies very close to the line of sight of 
  a dust lane, and also very close to the Wolf-Rayet star HD
 93162. (Source L has been scaled to match source E at high energies
 to see whether the difference in spectra can be accounted for 
 by extra absorption.) Fig. 10
 shows that while the spectral slopes of sources E and L match for energies
 higher than 1 keV, the softer flux from source L is much smaller than that of
 source E, which is consistent with the higher optical 
 extinction in source L.  (Massey and Johnson were particularly
 careful in the spectroscopic observations of the absorbed star because of
 their surprise at the early spectral type for a comparatively faint star.)
Source L is a luminosity class I star, where Source E is class V, but 
the major differences between the spectra seem to be due to different 
extinction.  Note that in Table 3, the column density appropriate to the
E(B-V) is between the standard cluster extinction and that found in a 
free fit.  By inference, the temperature would be between the temperatures
for the two solutions, which is comparable to the temperature found for 
Source E  and the standard cluster extinction.  This is consistent with 
the working hypothesis that the sources are a similar temperature, 
but with a different extinction.  

\subsection{O3f Star}  Fig. 11 shows the comparison of source F
(HD 93129AB) with the reference spectrum.  Source F is 
the brightest star in the Tr 14-16 region list of Massey and 
Johnson.  It is also the most unusual of the low-resolution 
X-ray spectra.   Fig. 11 shows prominent emission 
lines at the locations of ionized Mg, Si, and perhaps S.  This is 
in marked contrast to the smooth spectrum of source E at this resolution,
which is the second brightest star in the Massey and Johnson list.
The Helium-like  Mg XI and Si XIII lines are the strongest lines 
in the high-resolution  spectra of the O4f star $\zeta$ Pup
in this energy region (Cassinelli, et al., 2001; Kahn, et al., 2001).
As discussed in Sec. 2, HD 93129AB is a complex source (though 
unresolved in our image), consisting of components Aa, Ab, and B.  Since 
many of the other massive stars we detected are also multiple systems 
(Table 8, Paper I), the multiple components are not the cause of 
the unique spectrum in Fig. 11, though binarity may have some 
affect on the spectral properties.  The unique property of 
HD 93129Aa is that it is the most luminous star in the two 
clusters after $\eta$ Car.  The implication is that the spectral emission is 
related to the extreme luminosity, presumably coupled with a 
high mass loss rate and possibly circumstellar material.  

\subsection{Later O Stars} Fig. 12 shows the comparison of the later
O stars.  The spectra have not been scaled, so there is a range of count
rates.  However, the spectra are similar in that they all have 
soft counts.  Counts decrease with energy for energies greater 
than about 0.8 keV.  Fig. 10 in Paper I shows that there in no 
indication that L$_X$/ L$_{bol}$ is a function of photospheric 
spectral type.  This is consistent with Table 1 in this paper, 
which is ordered by optical magnitude.  The group of later O stars 
shows that neither the X-ray counts (column 6) nor the optical 
spectral types (column 12) is tightly correlated with the optical
magnitude.  Binarity for this group of X-ray sources (Table 8, Paper I)
is mixed.  Source B has no velocity evidence for binary 
motion, and the evidence for sources P and N needs to be confirmed.
The other three (O, M, and H) do show binary motion.  In other words, 
the level of X-ray strength is not uniquely dependent on optical 
luminosity, spectral type, or binarity.  The similarity of the 
X-ray spectra (Fig. 12) and  L$_X$/ L$_{bol}$ show that within the range 
of optical luminosity, spectral type, and multiplicity, the 
X-ray production is similar.  For the four sources that were 
fitted with temperatures (B, H, M, and N, Table 3), 
the constrained fit 
(requiring that the absorption be appropriate for the cluster)
found the temperatures for H, M. and N similar within the errors.
Source B has a somewhat cooler temperature from the fit.  However,
the similarity of the spectra in Fig. 12 underscores the fact that care 
must be taken in interpreting the fits when each bin has relatively
few counts.

\subsection{O3 vs O6 Stars} Fig. 13a shows the comparison between the reference
O3 spectrum (source E) and an O6 spectrum
(source B), scaled to match near 1 keV. The
slopes for E $\ge$ 0.8 keV are similar, however, the O6 star has
proportionately more counts than the O3 star at lower energies. The optical
absorption [E(B--V)] for the two stars is similar. In Fig. 13b, a second O3
star (source C) is compared with the O6 star. In this case, the spectra are
virtually identical. Thus, stars with optical spectral type O3 can have
different X-ray spectra (Fig. 9). On the other hand, stars with optical
spectral types O3 and O6 can have identical X-ray spectra (Fig 13b).
This is in accord with the hardness ratios in Paper I.  The three 
sources with particularly soft HR$_{MS}$
(soft hardness ratios) are 
sources G (HDE 303308), C (HD 93205), and B (-59$^o$2600). 
HR$_{MS}$ for source E, on the other hand, 
is harder, which agrees with Fig 13a.  One of the aims of this paper
is to explore how well the hardness ratios reflect the details 
of the spectra.  For these three sources, we can accurately 
differentiate between harder spectra (source E) and softer spectra
(B and C) using hardness ratios.  We note, however, that sources
E and C do not show orbital motion in their velocities (Table 8, Paper I).
Source C is a binary (O3 V + O8 V).  Does this invalidate the 
interpretation of the hardness ratios?  In cases where there is a 
colliding wind, the spectrum of the binary will be different from 
a non-interacting case, where  the composite spectrum is simply the 
sum of the spectra of the two stars.  Because L$_X$/ L$_{bol}$ is 
approximately constant for O stars, in the  case of a 
non-interacting composite spectrum, the X-ray 
spectrum will be dominated by the brightest star in the optical.
This situation would explain the similarity between the spectra of 
source B and source C in Fig. 13b.

\subsection{Source A: An O8.5 Star}  Fig. 14 shows the comparison between 
Source A, an O8.5 star and 
Source H, an O8 star, also shown in fig 12.  Neither spectrum has 
been scaled and Fig. 14 shows that at energies higher than 1 keV, 
Source A has about a factor of two more counts than source H.
On the other hand, if Source H were scaled to have the same flux 
as Source A for energies greater than 1 keV, Source A would have fewer 
soft counts than Source H.  This is    
in keeping with the extra absorption in Source A (Table 1).
Source H is a multiple system, with components O7 V, O8 V, and
O9 V (which is itself a binary; Albacete Colombo, et al. 2002).
Source A, on the other hand, has no information about radial 
velocity variation.  Fig. 12 shows that Source H has a spectrum
with a slope very similar to other late type stars, so the 
interpretation in the previous section that for a non-interacting
binary system, the spectrum will be dominated by the most 
luminous star appears to hold.  In contrast, the spectrum of 
source A is both harder and more absorbed than Source H (and 
the other late O stars in Fig. 12).  This would be consistent
with additional circumstellar material, possibly a colliding wind
system or an interacting binary.

\subsection{The Wolf-Rayet Star}  Fig. 15 shows the comparison 
between the spectra of the Wolf-Rayet star WR25 = HD93162 
(Source D) and the O3
standard star (Source E).   Fig 15a shows both spectra, unscaled
(source D: with error bars; source E: solid line). The 
spectra are strikingly similar and uncomplicated.  The dip near 2 keV
is an instrumental effect. In Fig 15b, 
the Wolf-Rayet spectrum (solid line) has 
been scaled to emphasize the similarity of the spectra at high
energies.  However at low energies, the Wolf-Rayet star has relatively
fewer counts than source E (dashed line), 
appropriate for the its higher absorption.    In general, WR stars 
have higher mass loss rates than O stars, which may account for
the lower count rate at low energies in the Wolf-Rayet star.
A major motivation for intercomparing the spectra of various 
sources in this series of plots is to investigate empirical characteristics 
of the spectra.  Experience with spectral fits to these comparatively
low resolution spectra shows that the fits are not unique. The fit in 
Table 2 for Source D with three components actually has a somewhat 
worse reduced $\chi^2$ than the simpler two component fit in Table 5.  
We have no particular expectation that the Wolf-Rayet star (Source D)
will have a spectrum similar to the O3 V star (source E), which has a
comparatively normal optical spectrum.  (We have no explanation, however, 
for the higher X-ray temperature of source E than other O3 V stars
in Fig. 9, particularly since there is no velocity evidence to 
support the possibility that source E might be a colliding wind binary.)
The spectral comparison in Fig. 15 supports the interpretation that 
the Wolf-Rayet star has a higher temperature X-ray component and larger
absorption than source E.  The fits to source D are discussed further
in Sec. 6.3, and compared with the high resolution XMM results of the 
same source (Raasen, et al. 2003).

\subsection{Surprisingly Hard Sources}
  Despite very different photospheric               
spectral types of source L (O3 I) and source A (O8.5 V), the X-ray spectra            
are identical, as shown in Fig. 16.  
Source A is also shown in Fig. 14, which illustrates
that it is an absorbed spectrum, confirming a large extinction in the
optical.  However, the spectra in Fig. 16 are not hard simply because 
their soft counts have been absorbed;   the count rate at 2 keV and higher
show that they are intrinsically hard.  In Table 3, most of the solutions 
listed  have constraints either on the temperature of the source 
or the absorption because of fits are not definitive when too many parameters
are included in a fit of weak spectra.  For source A, when both 
temperature and absorption are fitted, both are larger than the 
standard constraints for the cluster.  Although the temperature in 
the fit with constrained extinction 
(10.2 K) is not very hard, it is harder than constrained fits
for other O stars in Table 1, except for sources  D and F (and L) which are 
more evolved.   

\subsection{Cool Stars} 

Fig 17 shows the spectra of three stars (Sources I, J. and K) which are
 fainter  and cooler than the O star X-ray sources.  In Paper I, 
we overplotted all the O and B stars in the clusters on the X-ray 
sources.  Sources not coinciding with an O or B source are listed 
as cooler stars in Tables 2 and 4 in Paper I.  Sources I, J, and K
are clearly among the very strongest of these sources.  They have 
V magnitudes of 16.82, 14.34, and 13.16 mag (Table 1) respectively, appropriate 
for sources cooler than the O stars and fall within the 
 pre-main sequence band.  
For comparison, a representative
hot star spectrum (Source B, O6 V) is shown in Fig. 17.  All three cool stars
have fairly flat spectra. In contrast, the hot star
spectrum has proportionately more soft counts.  This is in 
accord with previous findings that cool stars produce X-rays
through magnetic processes at higher temperatures than the 
wind shock processes in hot stars.

\section{Discussion}

The goal of this investigation is to explore what physical X-ray
parameters can be determined from ACIS low-resolution spectra.
We have described a
particularly useful Chandra pointing for this purpose, because the 
region around $\eta$ Car contains a large number of  
extremely massive stars as well as many low mass stars.  

\subsection{Temperature Fits}

We have investigated the spectra  with detailed wind models,
with SHERPA spectral 
fits, and with comparisons between spectra.  The first step
is to discuss  whether the results 
from these approaches are consistent.  

As discussed above, the spectra of the weak sources do not constrain both the
extinction and the source temperature. For the stars in this field, 
 the extinction is surprisingly uniform, with 
 E(B-V) approximately 0.5 mag 
(MJ93). The hot stars typically have individually
determined reddenings (e.g. MJ93), which corresponds to a hydrogen
column density N$_H$ = 3 x 10$^{21}$ cm$^{-2}$ (Seward, 2000).
Table 3 lists a fit for each
source for which the ISM 
N$_H$ has been fixed to this value (``restricted fitting").
We will examine the temperatures using this fixed N$_H$ value to 
see if they are reasonable.  This value of absorption would be the 
minimum value, assuming there is no absorption in the wind
(the ``warm absorber").  From the temperature fits, we can see that 
this minimum absorption produces a high temperature.  If  the absorption
is allowed to increase in the fit, the fitted temperature decreases.
We stress that any fit is not unique;  by selecting the restricted 
fits for discussion, we have imposed a reasonable condition.

The cool stars (Sources I, J, and K) have harder 
spectra than the hot stars (Fig. 17).  In Table 3, two of the three cool 
stars, Sources I and K, have temperatures of 1.4 and 5.2 keV respectively
from this fitting, i.e. well above the standard range for hot stars. 
The other two sources for which  the standard 
E(B-V) is a poor choice are A and L, which we know 
from the optical results to have E(B-V) 
about twice that value.  Again this restricted 
fitting results in temperatures kT of 1.1 and 9.7 keV respectively,
 larger than the standard range for hot stars.  
For the remaining sources,  a combination of N$_H$ (consistent 
with optical reddening) and kT
can be found for a Raymond-Smith plasma that matches the spectrum.

 A recent paper by Albacete Colombo et al (2003) discusses XMM-Newton
 X-ray observations of many of the same targets as we do.  Because 
 our fitting procedures are somewhat different than theirs, the
 results from temperature fitting (their Table 4) will not be 
 identical to ours.  Our Table 3 presents fits generally restricting
 either the temperature or the absorption.  Similarly, our Table 5 
 presents fits for a few sources with two temperatures but a fixed 
 extinction. Since the foreground extinction  is 
 reasonably constant, so we have forced an extinction which is at 
 least a lower limit, rather than allowing what seem to be unphysically
 large values that sometimes result from allowing the extinction
 to be a free parameter.  
 
Table 6 shows the comparison between the single component fit in 
Table 3 and dominant component in the XMM fit.  
   Clearly, we match the shape of the fitted spectral function; the
differing bandpasses (XMM goes 0.2 keV softer) has a significant impact
on the precision and value of the fitted column density, particularly
for high values of the column density (sources B, C, D, E).  When the
two sets of fits yield approximately the same column, we obtain approximately
the same luminosity (sources A, G).  Remaining luminosity differences are
easily attributed to the wider XMM bandpass plus the poorly-calibrated phase
during which our data were obtained.

In general, our analysis of more complex multicomponent fits
agrees with that Albacete Colombo et al (2003)
in identifying sources with unusually hot components, which we 
will discuss in turn.
 For Source D = WR25, both analyses find one component a little above 
 3 keV and one close to 0.8 keV, although the absorption for the 
 cooler component is larger than our canonical absorption.  For
 source C = HD 93205 the temperatures in both fits are close to 
 1.0 keV and 0.2 keV, with approximately the same absorption.  
 For source E = HD 93250, the temperatures for both components are 
 hotter than our results (5 rather than 3 keV and 0.8 rather than
 0.3 keV), with similar absorptions.  For source A = CPD -59 2629,
 our Fig. 14 clearly shows that there is a hot component, and we know
  extra absorption is present from the optical results.  A hot component
 is found by  Albacete Colombo et al., as one of the components of the 
 fit.  Since we only did a one temperature fit, the derived temperature
 is lower than for the hottest component of their two temperature fit.  
 For source G = HDE 303308,  Albacete Colombo et al. found temperatures
 and extinctions which differ markedly for their two components.  For 
 our three fits, we found a single temperature and extinction 
 solution which was fairly close to standard values for the cluster
 for each.  For source B = CPD -59 2600, both our constrained fits 
 and the single temperature fit of  Albacete Colombo et al. found 
 temperatures and extinctions close to those typical for the cluster.  
 Direct comparisons of C and G (Fig. 9) and C and B (Fig. 13) 
 show that all three spectra are very similar, which agrees with
similar values from  our  temperature fits.

\subsection{Hardness Ratios}

In paper I we derived hardness ratios from 
 3 energy bands, 0.5 to 0.9 keV, 0.9 to 1.5 keV, and 1.5 to 2.04 keV,
 referred to as soft (S), medium (M), and hard (H) bands.            
Included  are the soft and hard hardness ratios:                  
 
$$ HR_{MS} = {(M-S)\over{M+S}} $$                                    
                                                                     
$$ HR_{HM} = {(H-M)\over{H+M}} $$                                    

We discuss HR$_{MS}$ but include 
HR$_{MH}$ in Table 1 for convenience. 
Fig. 18 in Paper I shows that HR$_{MS}$ and HR$_{HM}$
are correlated.  Hardness ratios are 
affected by reddening,  however, Fig. 18 in Paper I shows that 
the effect for a reasonable range of reddening is small compared 
to the observed range in the ratios.  

For the following discussion, for any source best fit 
with a two temperature model, we adopted the mean of the
fitted values.
In Fig. 18 the relation between $HR_{MS}$ and the temperature
from the ``restricted fitting" is shown. 
The general trend in Fig. 18 is the  same 
whether the temperature used for the brightest sources (D, E, and F) 
is the mean of the two temperature fit, or the temperature from the 
single temperature fit.  We stress that the temperature/absorption 
fits (especially for the weak sources) depends on which parameters
are fixed.  The stars in Fig. 18 show two very 
different behaviors.  The weaker stars have a range of hardness 
ratios, and a small to moderate range in temperatures.  The 
three strongest sources, however, have much larger 
temperatures and harder hardness ratios.  

Fig. 19 shows the temperatures as a function of photospheric
spectral type.  As in  Fig. 18, the weaker sources have 
a relatively small temperature range, which is not correlated 
with photospheric spectral type.  On the other hand, the  
4 stars with the hottest spectral types (O3 stars) divide between 
relatively cool and  much hotter X-ray 
temperatures.  In other words, the photospheric temperature
alone is not the determinant of the X-ray temperature.
The known binary properties of the sources
from radial velocity studies (Table 8, 
Paper I) provide a similarly mixed result. Of the two 
hot O3 sources, one is a binary, one is not.  For the cooler
O3 stars, one is binary, one  ``needs confirmation".  
Among the later spectral types, one does not have binary velocity 
variations, two are binaries, and 
one  ``needs confirmation".

Fig 20 shows the X-ray temperature as a function of the V 
magnitude.  For the cluster stars, this is related to the 
luminosity of the stars, and probably also to 
the mass loss rate.  The fainter stars (x's) show no relation between 
the X-ray temperature and the mass of the stars.  The X-ray 
temperatures are  markedly larger for the massive 
stars (squares).  
The correlation would be even more prominent except 
that the hottest star in X-rays is the Wolf-Rayet star, which  may well 
be somewhat faint in V because it has already lost mass.

\subsection{WR 25}

The components we found (Table 2) for WR 25 = HD 93162 (source D) can 
be compared with those found from fitting a high resolution XMM-Newton
spectrum (Raassen, et al. 2003).  
The combination of temperatures
and wind absorption values they find is somewhat different 
from those in Table 2, in that the XMM spectrum is fit
with two components, while the Chandra spectrum is fit 
with three. This makes detailed comparison difficult.
 Raassen et al. find  
 lines  seen in the XMM-Newton spectrum 
are from the same species as those in HD 93129AB (source F, Fig. 11). 
 The two features which are 
present between 1.0 and 2.0 keV in the XMM spectrum can 
be identified in the Chandra spectrum of HD 93162.  
The contrast in these
features, which is much smaller than those in HD 93129A,
is commensurate with that in the XMM spectrum.
The fits in Table 2 require much larger line widths for 
HD 93162 (source D) and HD 93250 (source E) than for 
HD 93129AB (source F).  Fig 3 shows that some of the width is 
accounted for by the distance  of the sources from the ACIS readout.
However, Raassen et al. list 2480 km sec$^{-1}$ as the terminal
velocity for WR 25.  This would contribute to the broadening
found in Table 2, but is only about 10\%\/ of the required 
amount.  The high velocities in WR 25 contribute to the  
reduction in the contrast in the lines.

Raassen et al. find a prominent Fe K shell feature near 6.5 keV.
While the XMM spectrum shows that this line is clearly present, 
it is not obvious in our spectrum (Fig. 15), presumably 
because it has relatively few counts.  
Because shock velocities in single stars are not usually strong
enough to produce this feature, they propose that it arises
in systems that are colliding wind binaries. The table they assemble
of Fe K$\alpha$ detections in binary and single stars, however,
shows mixed results on this question.

\subsection{X-ray Determinants}

The number of massive stars in Tr 14 and Tr 16 makes this field
a good place to attempt to identify factors contributing
to X-ray flux, particularly unusually strong X-ray flux.

The well-known relation X-ray luminosity and bolometric luminosity
 L$_X$ = 6 x 10$^{-7}$ L$_{bol}$ (see the recent 
 discussion in Albacete Colombo, et al 2003) implies a relation
with luminosity, and probably with mass loss and terminal velocity. The four
unusually strong, hard X-ray sources ($\eta$ Car itself, WR 25, HD 93129AB and
HD 93250) are all at the very end of their main sequence lifetimes or beyond. 
As discussed in Sec. 2 above, HD 93129AB is made up of at least three
components which are not resolved in our X-ray image, with spectral 
types for the A and B components of O2If* and O3.5 V((f+)) respectively.
The V magnitudes of Aa, Ab, and B are approximately 7.6, 8.5, and 8.9 
 mag respectively.  Based on the standard  relation between X-ray 
luminosity and bolometric luminosity, it would be expected that 
the brightest component Aa would also dominate in the X-ray region.  
In addition, Fig. 20 shows that stars of the magnitudes Ab and B 
would be expected to have a lower X-ray temperature than a star as 
bright as Aa.  While we cannot measure the individual X-ray 
brightnesses for the three components, it is likely that the 
X-ray spectrum is heavily dominated by the brightest star in the 
system, HD 93129 Aa.  
                                    
The X-ray properties of $\eta$ Car itself are set out by Albacete Colombo et
al. It has an unusually large X-ray luminosity and X-ray to bolometric
luminosity ratio. They fit the X-ray spectrum with four Raymond-Smith
components with temperatures up to 4.4 keV. It also has a strong Fe K $\alpha$
line. 

Stellar radius increases in very late-stage main sequence stars, yielding
increased mass loss. This characteristic should affect $\eta$ Car itself, WR
25, HD 93129AB and HD 93250. Clearly there are several possible variants to
this scenario, in particular a binary system (Paper I), 
or a confined magnetic field.  As discussed
carefully by Raassen et al., information on
the binarity of WR 25 is inconclusive. For $\eta$
Car and HD 93129AB, binary effects may certainly be important. HD 93250 has no
velocity evidence of a binary companion. Fig. 10 in Albacete Colombo et al.
shows that hard spectra are not correlated with binary companions.

 Schulz et al. (2003) suggest that $\theta^1$ Ori C, A, and E all
are magnetic stars  because they are very close to the
zero age main sequence. This adds another possibility to the list of complexities.
One candidate source for confined magnetic fields is source A = Tr16-22. It
has an unusually high L$_X$/L$_{bol}$, as well as a surprisingly hard
spectrum, particularly for an O8.5 star. If this is the case, it would be an
example of a peculiar property determining the X-ray production. It will be
very interesting to see what the radial velocity studies undertaken by
Albacete Colombo et al. (2003) reveal.  Albacete Colombo et al. do find that
it is not variable in their 43 ksec  XMM observation.  In other words
they do not see a flare which would be a clue to X-rays produced 
by a pre-main sequence companion.

If it turns out that magnetic fields play a role for a 
significant fraction of O stars, 
it may even be that a close binary companion can suppress
a magnetic field, thus actually reducing the X-ray flux.

\subsection{Evolutionary Development}

Qualitatively, the comparison of the X-ray spectra of WR 25, HD 93129AB, and  
HD 93250 shows dramatic differences.  Is the prominence of emission 
features  related  simply to evolutionary changes?  Certainly 
decreasing in mass loss rate in this sequence causes the X-ray 
spectrum to sample different levels of the atmosphere.  The E(B-V)
of WR 25 is larger than for HD 93129AB and HD 93250, as would be 
expected from the high mass loss rates of WR stars.  However, the 
X-ray spectrum of the Wolf-Rayet star $\gamma^2$ Vel provides a 
cautionary note about a simple interpretation (Skinner, et al. 2001;
Willis et al. 1995; Stevens et al. 1996).  Willis, et al. (1995) and 
Stevens, et al. (1996) obtained ROSAT and ASCA observations of 
$\gamma^2$ Vel respectively.  With the ROSAT observations, they 
mapped a dramatic increase in X-ray flux close to the periastron 
in the 78$^d$ orbital period.  The ASCA observations also show a 
dramatic change in two spectra before and after periastron.
After periastron, the X-ray flux was both larger and harder, 
and also has emission lines very similar to those seen in 
HD 93129AB.  This was confirmed by a Chandra HETG spectrum 
at the same phase (Skinner et al. 2001).  
The proposed explanation is that the 
large X-ray flux appears when the system is viewed along the shock cone of 
the colliding wind (around the O star, which has a weaker wind
than the WR star).  At other phases, the extended WR star 
dominates the spectrum.  Thus the "wind-shock" phase in 
$\gamma^2$ Vel produces a spectrum much more like the O3 If 
HD93129AB than WR 25.  Even the spectrum in $\gamma^2$ Vel
at the phase dominated by the Wolf-Rayet star is softer than 
WR 25.  In other words, sources may be made up of several 
components, and it is only the relative importance which changes 
as the stars evolve.  
Furthermore, as noted above, emission lines are seen in the 
XMM spectrum of WR 25 (Raasen, et al 2003).  The addition of 
a hard component to a spectrum like HD 93129AB could swamp 
the emission features.  

In the previous section we have discussed properties
such as binarity and magnetic fields which 
may mean that the evolutionary development may differ 
from evolution of a single object.  Ignoring these complexities,
we suggest here a simple scenario which 
relates the changes in the X-ray spectra to evolutionary
changes in these very massive stars.  First it is clear 
that HD 93205 (= source C) should become a colliding wind binary in the future
since it is in a short period (6$^d$) orbit with an O8 V 
star.  However, so far its spectrum resembles cooler
 O stars, and  the X-ray bright
O3 V star HD 93250  ( = source E; e.g. Figs 9 and 19).  HD 93250  
 is more X-ray luminous, as well as more bolometrically luminous.
The mass loss rates of the sequence of O3 stars HD 93205,
HD 93250, and HD 93129AB (= source F) are log  \.{M} =
  -5.8, -4.6, and -4.9 M$\sun$ yr$^{-1}$ respectively
(Howarth and Prinja, 1989) from IUE high resolution spectra.
Thus it is likely that the increase in X-ray luminosity from 
HD 93205 to HD 93250 is related to the increased mass loss 
in the more luminous, and presumably more massive star.  
The O3 If supergiant system HD 93129AB is not only more  
X-ray luminous, but it also has very broad emission lines.  
Circumstellar material accumulated from the mass loss of the 
more evolved primary star in the system could account for 
the emission, even though the mass loss rate found by 
Howarth and Prinja is actually smaller than that for HD 93250.  
For HD 93162 = WR 25 = source D, the mass loss rate log  \.{M}  -4.4 
 M$\sun$ yr$^{-1}$ (Raasen, et al. 2003).  It is an unusually 
 strong X-ray source  compared with other WR stars. 
 The strong hard X-rays, the large mass loss rate, and the 
large E(B-V) are all presumably related.  The $\gamma^2$ Vel system 
demonstrates that a WR system can produce broad emission lines,
although they are related to the wind-shock cone.  It is not clear 
whether WR 25 is a binary system, as discussed by Raasen et al., 
although they find a Fe K$\alpha$ feature which they suggest 
signifies a binary system.  As suggested above, in our scenario 
emission lines related to circumstellar material such as those 
in HD 93129AB may be dominated by the unusually strong high 
temperature component in WR 25.  Although this scenario for the 
four strongest X-ray sources is 
simplified, it does provide a working model which links  the 
observed features in the X-ray spectra with the underlying 
evolutionary changes.  

\section{Summary}

Among the points found from the spectral comparisons in this 
study are:

$\bullet$ Among the hottest (O3) stars with similar photospheric
temperatures (Fig. 9), two have identical
spectra, but one is significantly harder.  The harder spectrum is 
produced by the more bolometrically
luminous star.  A harder spectrum in a more 
luminous star may result partly from
extra  absorption from a larger mass loss in the wind.  However, it
also has a higher X-ray flux, which means the difference cannot 
result only from the removal of soft counts.  

$\bullet$  Extra absorption is clearly seen in the spectrum of 
source L = Tr 16-244 ( Fig. 10).  This may be from foreground absorption
from the nearby dust lane, circumstellar absorption, or 
absorption related to the proximity to the Wolf-Rayet star WR 25 (HD 93162). 

$\bullet$  The spectrum of the supergiant HD 93129AB shows 
pronounced emission lines, even in comparison with the 
Wolf-Rayet star (Fig 11).

$\bullet$ Later O stars have spectra similar to the O3 main 
sequence stars (Figs 12 and 13).

$\bullet$  The comparison (Fig 14) of the spectra of the heavily reddened 
 O 8.5 star source A   with source H 
 (normal extinction) for the region shows that  the  reddened 
star also has a harder spectrum with an unusually large number of counts
at 2 keV for an O star.

$\bullet$   The Wolf-Rayet star source D (Fig 15) has a spectrum very 
similar to the luminous O3 star source E (HD 93250)
 in that it has significant 
number of hard counts, and no particularly prominent lines at
this resolution.  The high extinction of the Wolf-Rayet 
star is apparent from the comparison of the counts below 1 keV.

$\bullet$   The spectra of two sources known to have high 
absorption (source A O8.5 V and source L O3 I, Fig 16) also have 
high counts at 2 keV, showing that they are intrinsically 
hard.  

$\bullet$ The spectra of 3 cool stars (Fig 17) are harder 
than a typical hot star, even though the cool star spectra 
are relatively weak.

$\bullet$ The hardness ration  $HR_{MS}$ is correlated with the 
X-ray temperature.  (Fig. 18).

$\bullet$ The X-ray temperature is not correlated with the photospheric
spectral type (Fig. 19).

$\bullet$ The strongest determinant of high X-ray temperature is the 
high luminosity of the star (Fig. 20), although other factors
such as a binary companion or a magnetic field may have an 
effect.

Finally, we return to the three goals in the introduction which summarize 
the properties of the X-ray bright stars.  

$\bullet$ What happens to the X-ray properties as the very massive
stars evolve?  Qualitatively, the sequence HD 93250 to HD 93129AB
to HD 93162 = WR 25 is a sequence of very massive 
stars from the main sequence through the first stages of the 
Wolf-Rayet phase toward $\eta$ Car.  The X-ray spectra 
for these stars are all hard, but with dramatic changes, 
from relatively smooth through prominent emission lines 
(HD 93129AB) back to comparatively smooth for WR 25. 
Presumably this results from increasing mass loss driven by 
evolution. Different regions would then dominate 
the spectra because of different optical depths.  However, there  
may be complicating factors governing the X-ray spectra 
such as binary interactions (colliding winds) or magnetic 
fields.  

$\bullet$ How are the stellar photospheric parameters and 
absorption related the the X-ray spectra? Sources A and L
which have high absorption  in the optical have extra 
absorption of soft X-ray counts, as would be expected.
The X-ray spectra of the hot stars can be divided into 
two groups.  For normal main sequence stars from O3 to O8,
the X-ray temperatures have no dependence on spectral type 
or luminosity.  For more luminous stars discussed in the 
immediately preceding bullet, have higher X-ray temperatures.
This distinction is picked up by the hardness ratios also.

$\bullet$ What are the X-ray characteristics of the cool stars?
Although we only have very weak spectra of a few cool stars, 
they are different (harder) than the spectra of the hot stars.  

Acknowledgments

Financial assistance was provided from the                                           
Chandra X-ray  Center NASA Contract NAS8-39073 for NRE, EMS, FDS, MIK, JN, 
and SJW.  WLW was supported in part by Chandra grants GO2-3027A and GO2-3028.
We thank the anonymous referee for very thorough comments which 
considerably improved the presentation and discussion of the paper, 
and a useful conversation with Fred Bruhweiler.

\clearpage

\begin{deluxetable}{llrrrrrrrrrlr}
\footnotesize
\tabletypesize{\scriptsize}
\tablecaption{ X-Ray Sources
\label{tbl-1}}
\tablewidth{0pt}
\tablehead{
\colhead{Src} & \colhead{ ID } & \colhead{ Num } &
\colhead{RA} & \colhead{ Dec} &  \colhead{Cts}& \colhead{HR$_{MS}$} & \colhead{
Err } 
& \colhead{HR$_{HM}$}& \colhead{Err}& \colhead{ V }& \colhead{Spect.}& 
 \colhead{ E(B-V) } \\
\colhead{} & \colhead{ } & \colhead{ } &
\colhead{} & \colhead{ } &  \colhead{}& \colhead{} & \colhead{ } &
 \colhead{}& \colhead{}& \colhead{ mag}& \colhead{Type}& 
\colhead{ mag } \\
}

\startdata


   F  & HD93129AB & 103 & 10 43 57.51 & -59 32 53.1  & 2132.4 &   0.45  &   0.03  &  -0.49  &   0.03  & 6.90 & O3 I & 0.55 \\
   E   &   HD93250 & 22 & 10 44 45.06 & -59 33 55.2 & 1572.8  &   0.13  &   0.03  &  -0.58  &   0.04  & 7.41 & O3 V & 0.49 \\
   C    &  HD93205 & 25 & 10 44 33.80 & -59 44 15.6 & 532.6  &  -0.24  &   0.05  &  -0.72  &   0.09  & 7.76 & O3 V & 0.40 \\
 D  & HD93162 & 28 & 10 44 10.44 & -59    43 11.2 & 4989.9  &   0.61  &   0.02  &  -0.42  &   0.02  & 8.11 & WN7 & 0.66 \\
  G    & HDE303308 & 10 & 10 45 05.95 & -59    40 06.3 & 232.1  &  -0.31  &   0.08  &  -0.63  &  0.14  & 8.19 & O3 V & 0.46 \\
  B    & -59$^0$2600 & 88 & 10 44 41.87 & -59    46 56.5 & 179.9  &  -0.18  &   0.09  &  -0.82  &   0.15  & 8.65 & O6 V & 0.51 \\
  O    & -59$^0$2603 & 83 & 10 44 47.36 & -59    43 53.4 &  52.2   &  -0.06  &   0.16  &  -0.72  &   0.23  & 8.82 & O7 V & 0.46 \\
   M   & -59$^0$2635 & 44 & 10 45 12.78 & -59    44 46.4 &  61.3  &  -0.06  &   0.16  &  -0.72  &   0.23  & 9.27 & O8.5 V & 0.54   \\
   P   & -59$^0$2641 & 2 & 10 45 16.58 & -59    43 37.3 & 84.7  &  -0.12  &   0.14  &  -0.65  &   0.20  & 9.28 & O6 V & 0.61 \\ 
  H   & -59$^0$2636 & 5 & 10 45 12.93 & -59 44    19.4 &  114.0 &    0.05  &   0.12  &  -0.77  &   0.17 & 9.31 & O8 V & 0.60 \\ 
  N   & HD93343 & 45 & 10 45 12.28 & -59 45 00.6  & 64.4  &  -0.03  &   0.16  &  -0.64  &   0.22  & 9.60 & O7 V & 0.56 \\ 
   L  & Tr16-244 & 102 & 10 44 13.26 & -59 43 10.4 &  142.7  &   0.66  &   0.13  &  -0.37  &   0.11   &   10.78 & O3 I & 0.94 \\
   A   & Tr16-22 & 55 & 10 45 08.26 & -59 46 07.3 & 226.4  &   0.51  &   0.10  &  -0.43  &   0.09  & 10.93  & O8.5 V & 0.78 \\  
    
 
  I   & & 71 & 10 44 57.96 & -59 47 9.5 & 51.3 & 0.82 & 0.27 & -0.06 & 0.17 &
    16.82  & &  \\ 
  J   & & 70 & 10 45 0.99 & -59 45 15.6 & 50.4 & -0.16 & 0.18 &
    -0.72 & 0.27 & 14.34 & &  \\ 
  K   & &  85 & 10 44 46.54 & -59 34 12.4 & 90.8
    & 0.44 & 0.16 & -0.38 & 0.15 & 13.16 & & \\

\enddata

\end{deluxetable}

\begin{deluxetable}{lllll}
\footnotesize
\tabletypesize{\scriptsize}
\tablecaption{ Table of Best-Fit Wind Model Parameters
 \label{tbl-1}}   
\tablewidth{0pt}
\tablewidth{6truein}
\tablehead{
\colhead{Component } & 
\colhead{Parameter } & \colhead{HD93129AB } & \colhead{HD93250 }
& \colhead{HD93162 }  \\
  }

\startdata

 &  N$_{ISM}$  /10$^{21}$  (cm$^{-2}$) & 2.9 & 2.6 & 3.4  \\

1st & T$_X$  (MK) & 5.6$\pm$ 3.2 & 6.4 $\pm$ 0.9 & 7.8 $\pm$ 4.1 \\
    & kT (keV)    &  0.48          & 0.55            & 0.67            \\

 &  EM /10$^{56}$ (cm$^{-3}$) & 2.1 $\pm$ 1.9 & 3.2 $\pm
 $ 1.0 & 4.4 $\pm$
4.0 \\ 
&  N$_W$ /10$^{21}$ (cm$^{-2}$) & $<$ 10 & 5.6 $\pm$ 1.5 & $<$ 12 \\ 
& F$_O$/10$^{-13}$ (erg cm$^{-2}$ s$^{-1}$) & 0.87 $\pm$ 0.10 & 7.1 $\pm$ 0.3 & 3.8 $\pm$
0.6 \\

 &F /10$^{-13}$  (erg cm$^{-2}$ s$^{-1}$) & 3.5  $\pm$ 0.4 & 60.4 $\pm$ 2.4 &
16.6 $\pm$ 2.6 \\

2nd & T$_X$  (MK) & 8.2  $\pm$  1.8 & 27.2  $\pm$  16.0 & 14.1  $\pm$  3.5 \\
    & kT (keV)    &  0.71          & 2.33            & 1.21            \\

 & EM /10$^{56}$ (cm$^{-3}$) & 3.7 $\pm$  1.2 & 5.6 $\pm$  4.3 &
9.8 $\pm$  3.1 \\
& N$_W$ /10$^{21}$ (cm$^{-2}$) & 10.6 $\pm$ 3.3 & 75.0 $\pm$ 47.8 &
7.6 $\pm$ 3.2 \\
 & F$_O$/10$^{-13}$ (erg cm$^{-2}$ s$^{-1}$) & 4.2 $\pm$ 0.5 &
2.4 $\pm$ 0.4 & 24.3 $\pm$ 3.8 \\
 & F /10$^{-13}$  (erg cm$^{-2}$ s$^{-1}$) & 58. $\pm$ 7. &
38. $\pm$ 6. & 120. $\pm$ 19. \\
3rd &  T$_X$  (MK) & $>$ 10 & -- & $>$ 7 \\
    & kT (keV)    &  $>$ 0.9          & --            & $>$  0.6           \\

 & EM /10$^{56}$ (cm$^{-3}$) & 0.9 $\pm$ 0.6 & -- & 13.9 $\pm$ 13.1 \\
& N$_W$ /10$^{21}$ (cm$^{-2}$) & $>$ 20 & -- &  $>$ 25  \\
& F$_O$/10$^{-13}$ (erg cm$^{-2}$ s$^{-1}$) & 0.8 $\pm$ 0.1 &
-- & 6.1 $\pm$ 1.0 \\
& F /10$^{-13}$  (erg cm$^{-2}$ s$^{-1}$) & 2.6 $\pm$ 0.4 &
-- & 250. $\pm$ 39. \\
& & & & \\
& $\chi^2$ / DOF & 97.3 / 81 & 144.4 / 116 & 849.3 / 224 \\
& Energy Shift (eV) & 29. & -140. & -56. \\
&HWHM (eV) & 7.0 & 147. & 126. \\
& Total F$_O$/10$^{-13}$ (erg cm$^{-2}$ s$^{-1}$) &
5.9 $\pm$ 0.6 & 9.4 $\pm$ 0.9 & 34.3 $\pm$ 5.4 \\ 
 & Total F /10$^{-13}$  (erg cm$^{-2}$ s$^{-1}$) &
65. $\pm$ 7. &  99. $\pm$ 10. & 387. $\pm$ 61. \\

\enddata                                                                       
                                                                                
\end{deluxetable}

\begin{table}
\begin{center}
\caption{Spectral Model Parameters Including the Weaker Sources\tablenotemark{a}}
\label{spec_mod}
\begin{tabular}{lrrrrrrrrr}
       &  &  &             &                &     &
\multicolumn{2}{c}{Absorbed Flux\tablenotemark{c}} &
\multicolumn{2}{c}{Unabsorbed Flux\tablenotemark{c}} \\
Source & N$_{\rm H}$\tablenotemark{b} & kT\tablenotemark{b} & T\tablenotemark{b}  
& ${\chi}^2$/dof & dof & 0.5-2.0 & 2.0-10 &
0.5-2.0 & 2-10 \\ \hline
A & 1.08$\pm$0.15 & 0.59$\pm$0.11 & 6.9 & 2.4 & 32 & 8.8$\pm$0.5e-14 & 2.3$\pm$0.3e-14 & 1.6e-12 & 3.0e-14 \\
  & 0.3f & 0.88$\pm$0.05 & 10.2 & 4.9 & 33 & 7.2$\pm$0.5e-14 & 2.5$\pm$0.3e-14 & 1.4e-13 & 2.7e-14 \\
  & 1.27$\pm$0.12 & 0.40f& 4.7 & 2.4 & 33 & 8.6$\pm$0.7e-14 & 1.3$\pm$0.2e-14 & 3.6e-12 & 1.9e-14 \\

B & 0.3f & 0.40$\pm$0.05 & 4.6 & 3.4 & 18 & 6.7$\pm$0.8e-14 & 3.9$\pm$1.0e-15 & 1.8e-13 & 4.3e-15 \\
  & 0.54$\pm$0.05 & 0.40f& 4.7 & 2.9 & 18 & 6.1$\pm$0.4e-14 & 2.1$\pm$0.3e-15 & 4.7e-13 & 2.4e-15 \\

C & 0.60$\pm$0.05 & 0.24$\pm$0.03 & 2.7 & 2.7 & 50 & 2.3$\pm$0.4e-13 & 1.5$\pm$0.5e-16 & 6.1e-13 & 4.8e-15 \\
  & 0.3f & 0.30$\pm$0.04 & 3.5 & 6.2 & 51 & $\cdots$ & $\cdots$ & $\cdots$ & $\cdots$\\
  & 0.44$\pm$0.06 & 0.40f& 4.7 & 6.9 & 51 & $\cdots$ & $\cdots$ & $\cdots$ & $\cdots$\\

D & 1.20$\pm$0.03 & 0.79$\pm$0.08 & 9.2 & 5.2 & 221 & 1.9$\pm$0.2e-12 & 8.8$\pm$0.9e-13 & 7.8e-11 & 1.3e-12\\
  & 0.3f & 2.72$\pm$0.12 & 31.5 & 8.2 & 222 & $\cdots$ & $\cdots$ &  $\cdots$ \\
  & 1.44$\pm$0.10 & 0.4f & 4.7 & 16.3 & 222 & $\cdots$ & $\cdots$ &  $\cdots$ \\ 

E & 0.76$\pm$0.04 & 0.65$\pm$0.22 & 4.9 & 7.6 & 112 & 5.8$\pm$0.6e-13 & 1.1$\pm$0.7e-13 & 5.2e-12 & 1.3e-13\\
  & 0.3f & 0.83$\pm$0.02 & 9.3 & 8.6 & 113 & $\cdots$ & $\cdots$ &  $\cdots$ \\
  & 0.95$\pm$0.03 & 0.4f & 4.7 & 7.7 & 113 & $\cdots$ & $\cdots$ &  $\cdots$ \\ 

F & 0.95$\pm$0.05 & 0.80$\pm$0.03 & 9.3 & 4.4 & 105 & 4.9$\pm$0.6e-13 & 2.1$\pm$0.2e-13 & 8.8e-12 & 2.7e-13\\
  & 0.3f & 1.70$\pm$0.05 & 19.7 & 4.5 & 106 & $\cdots$ & $\cdots$ &  $\cdots$ \\ 
  & 1.32$\pm$0.04 & 0.4f & 4.7 & 8.4 & 106 & $\cdots$ & $\cdots$ &  $\cdots$ \\ 

G & 0.59$\pm$0.06 & 0.19$\pm$0.05 & 2.2 & 3.4 & 24 & 9.8$\pm$1.0e-14 & 2.8$\pm$0.8e-16 & 2.6e-12 & 2.8e-16 \\
  & 0.3f & 0.29$\pm$0.02 & 3.3 & 5.3 & 25 & $\cdots$ & $\cdots$ & $\cdots$ & $\cdots$ \\
  & 0.29$\pm$0.05 & 0.40f& 4.7 & 6.1 & 25 & $\cdots$ & $\cdots$ & $\cdots$ & $\cdots$ \\

H & 0.3f & 0.74$\pm$0.08 & 8.6 & 3.8 & 13 & 3.5$\pm$0.7e-14 & 3.2$\pm$0.5e-15 & 8.8e-14 & 3.4e-15 \\
  & 0.71$\pm$0.10 & 0.40f& 4.7 & 4.9 & 13 & 4.5$\pm$0.3e-14 & 2.4$\pm$0.7e-15 & 5.6e-13 & 3.0e-15 \\

I & 0.3f & 1.4$\pm$0.2  & 16.3 & 5.2 &  9 & 1.6$\pm$0.3e-14 & 9.8$\pm$0.6e-15 & 3.1e-14 & 1.1e-14 \\
  & 1.35$\pm$0.09 & 0.40f& 4.7 & 4.7 &  9 & 2.1$\pm$0.4e-14 & 3.6$\pm$0.3e-15 & 1.0e-12 & 5.5e-15 \\

J & 0.3f & 0.51$\pm$0.11 & 5.9 & 1.6 &  6 & 2.4$\pm$0.9e-14 & 7.3$\pm$1.1e-16 & 7.1e-14 & 7.9e-16 \\
  & 0.43$\pm$0.05 & 0.40f& 4.7 & 1.3 &  6 & 2.6$\pm$0.7e-14 & 6.3$\pm$2.0e-16 & 1.4e-14 & 7.2e-16 \\

K & 0.3f & 5.2$\pm$1.3 & 60.5 & 0.8  &  7 & 3.2$\pm$0.5e-14 & 1.0$\pm$0.4e-14 & 1.1e-13 & 1.1e-13 \\
  & 1.34$\pm$0.10 & 0.40f& 4.7 & 1.2 &  7 & 3.0$\pm$0.7e-14 & 5.3$\pm$1.0e-15 & 1.5e-12 & 7.9e-15 \\

L & 0.3f & 1.80$\pm$0.20  & 21. & 1.4 & 15 & 5.5$\pm$1.1e-14 & 2.5$\pm$0.5e-14 & 1.0e-13 & 2.6e-13 \\
  & 1.58$\pm$0.05 & 0.40f& 4.7 & 2.1 & 15 & 5.2$\pm$0.4e-14 & 1.2e-14$\pm$0.3 & 3.5e-12 & 1.8e-14 \\

M & 0.3f & 0.63$\pm$0.21 & 7.3 & 2.7 & 5 & 2.7$\pm$0.5e-14 & 1.5$\pm$1.0e-15 & 7.3e-14 & 1.6e-15 \\
  & 0.56$\pm$0.05 & 0.40f& 4.7 & 1.6 & 5 & 2.4$\pm$0.2e-14 & 8.4$\pm$0.4e-16 & 1.8e-13 & 9.8e-16 \\

N & 0.3f & 0.85$\pm$0.05 & 9.9 & 3.8 &  6 & 2.2$\pm$0.3e-14 & 3.3$\pm$0.7e-15 & 5.2e-14 & 3.6e-15 \\
  & 0.89$\pm$0.05 & 0.40f& 4.7 & 3.9 &  6 & 2.4$\pm$0.2e-14 & 1.9$\pm$0.3e-15 & 4.5e-13 & 2.4e-15 \\ \hline
\end{tabular}
\tablenotetext{a}{All spectral fits used Raymond-Smith model; f =
parameter fixed with a linear gain shift applied.}
\tablenotetext{b}{Units = 10$^{22}$ cm$^{-2}$ for N$_{\rm H}$; keV for
kT; 10$^6$ degrees K for T.}
\tablenotetext{c}{Units = erg s$^{-1}$ cm$^{-2}$ in the listed energy band.}
\end{center}
\end{table}

\begin{table}
\begin{center}
\caption{Model Fit to Source C}
\label{C_spec}
\begin{tabular}{llrrrr}
          &              & Parameter      & EqW or & \\
Component & Description  & Value          & Flux\tablenotemark{a} &
${\chi}^2$/dof, dof & ${\Delta}{\chi}^2$ \\ \hline
absorption & N$_{\rm H}$ & 0.34$\pm$0.05  & $\cdots$ & $\cdots$ & $\cdots$ \\
continuum  & R-S kT      & 0.37$\pm$0.05  & 4.4 $\times$ 10$^{-13}$ & 2.72, 49 & $\cdots$ \\
gaussian-1 & line E      & 0.570$\pm$0.008 & 153      & 2.42, 47 & 19.7 \\
gaussian-2 & line E      & 0.892$\pm$0.035 &  60    & 2.14, 45 & 17.4 \\
gaussian-3 & line E      & 0.686$\pm$0.010 &  65      & 1.76, 43 &
20.6 \\ \hline
\end{tabular}
\tablenotetext{a}{unabsorbed Flux in erg s$^{-1}$ cm$^{-2}$ 
 in 0.5-8 keV band; EqW = equivalent width in eV.}
\end{center}
\end{table}

\begin{table}
\begin{center}
\caption{Model Fit for 2 Temperatures}
\label{C_spec}
\begin{tabular}{llrrrrrr}
  
Source  & kT    &  kT$_1$ & T$_1$ &  kT$_2$ & T$_2$ & ${\chi}^2$/dof & dof  \\
       & (mean)   &    (keV) &  (MK) & (keV) &   (MK)
 &  &  \\  \hline
     
 C &  0.62 & 1.06 & 12.3 & 0.19 & 2.2 & 2.39 & 46 \\

D &  2.08 & 3.29 & 38.4  & 0.86  & 10.0   & 3.23  &  217 \\

E &  1.59 & 2.85 & 33.0 & 0.33 &  3.9 & 2.36 & 109 \\

F & 1.72 & 2.59 & 30.1  & 0.84  & 9.8   & 2.50  &  103 \\


 \\ \hline

\end{tabular}

\end{center}
\end{table}

\begin{table}
\begin{center}
\caption{XMM-Chandra Comparisons}
\label{C_spec}
\begin{tabular}{lllll}

Source & Satellite  &  N$_H$    &  kT & Flux   \\
       &           & 10$^{22}$ cm$^{-2}$  & (keV) &  10$^{-13}$ ergs s$^{-1}$ cm$^{-2}$ \\  \hline

 & & & & \\ 
A &  XMM &  1.2$\pm$0.1 &   0.37$\pm$0.07 &  30$\pm$1  \\ 
 &  Chan & 1.08$\pm$0.15 &  0.59$\pm$0.15 &  16 \\ 
 & & & & \\
B & XMM  & 0.4$\pm$0.4  &  0.30$\pm$0.01 &  8.8$\pm$0.3  \\ 
  &  Chan & 0.3f       &  0.40$\pm$0.05 &  1.8 \\ 
 & & & & \\ 
C & XMM  & 0.45$\pm$0.06 & 0.25$\pm$0.01 & 35$\pm$2 \\ 
  & Chan & 0.60$\pm$0.05 & 0.24$\pm$0.03 &  6 \\ 
 & & & & \\ 
D & XMM  & 0.81$\pm$0.02 & 0.76$\pm$0.08 & 256 \\ 
  & Chan & 1.20$\pm$0.03 & 0.79$\pm$0.08 &  19 \\ 
 & & & & \\ 
E & XMM  & 0.26$\pm$0.09 & 0.81$\pm$0.02 & 34$\pm$2 \\ 
  & Chan & 0.76$\pm$0.04 & 0.65$\pm$0.22 &  6 \\ 
 & & & & \\ 
G & XMM  & 0.56$\pm$0.1  & 0.23$\pm$0.04 & 25$\pm$2 \\ 
  & Chan & 0.59$\pm$0.16 & 0.19$\pm$0.05 & 26 \\ 
 & & & & \\ 
 \\ \hline

\end{tabular}

\end{center}
\end{table}

\clearpage

\figcaption[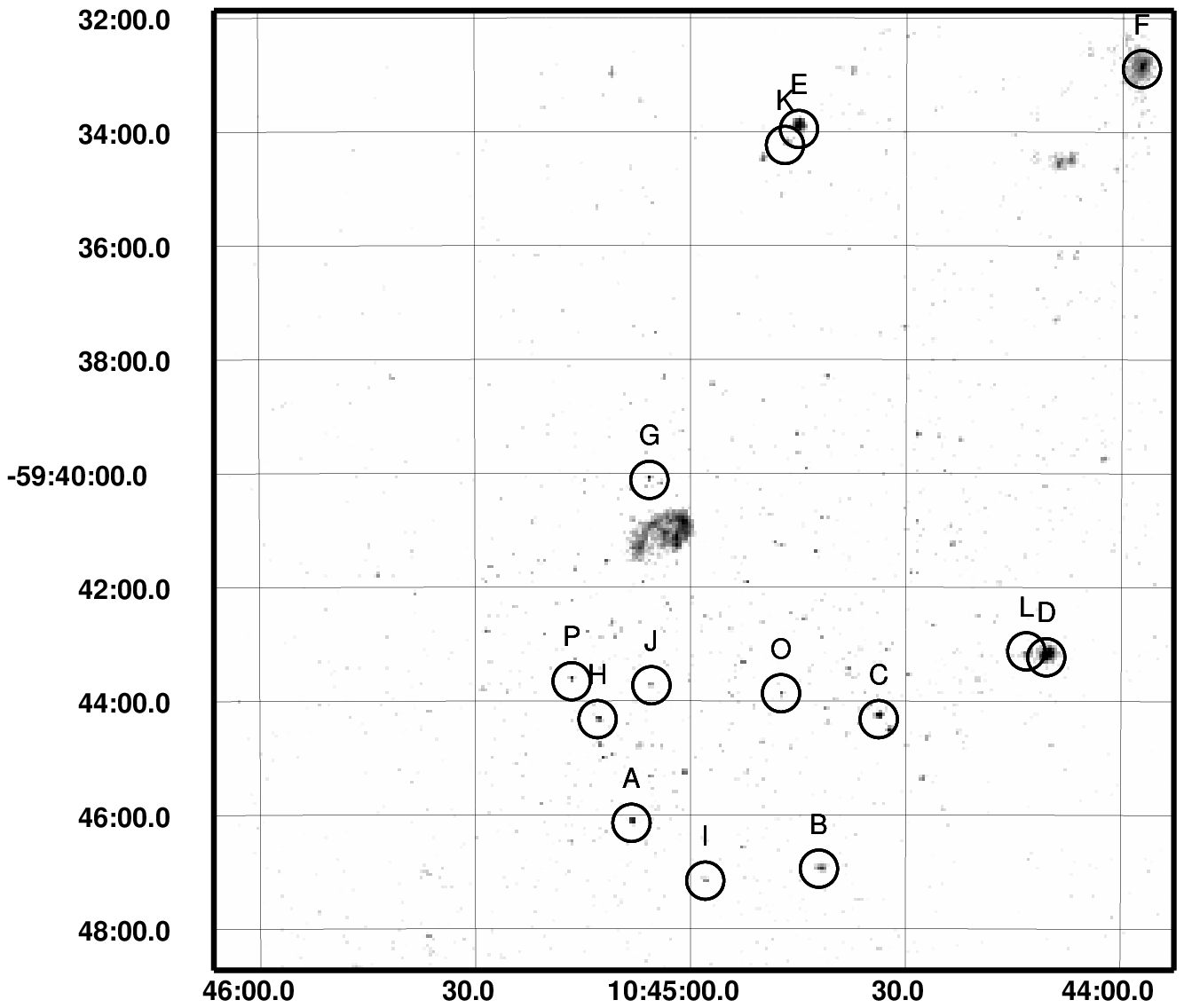]{The ACIS-I image of the region surrounding
$\eta$ Car. $\eta$ and the stars to the South are in Tr 16.
  Tr 14 is in the extreme upper right corner.  The positions
of the spectra in Table 1 are shown by the circles.  Sources        
m and n are the two sources immediately south of source h.     
   \label{fig1}}

\figcaption[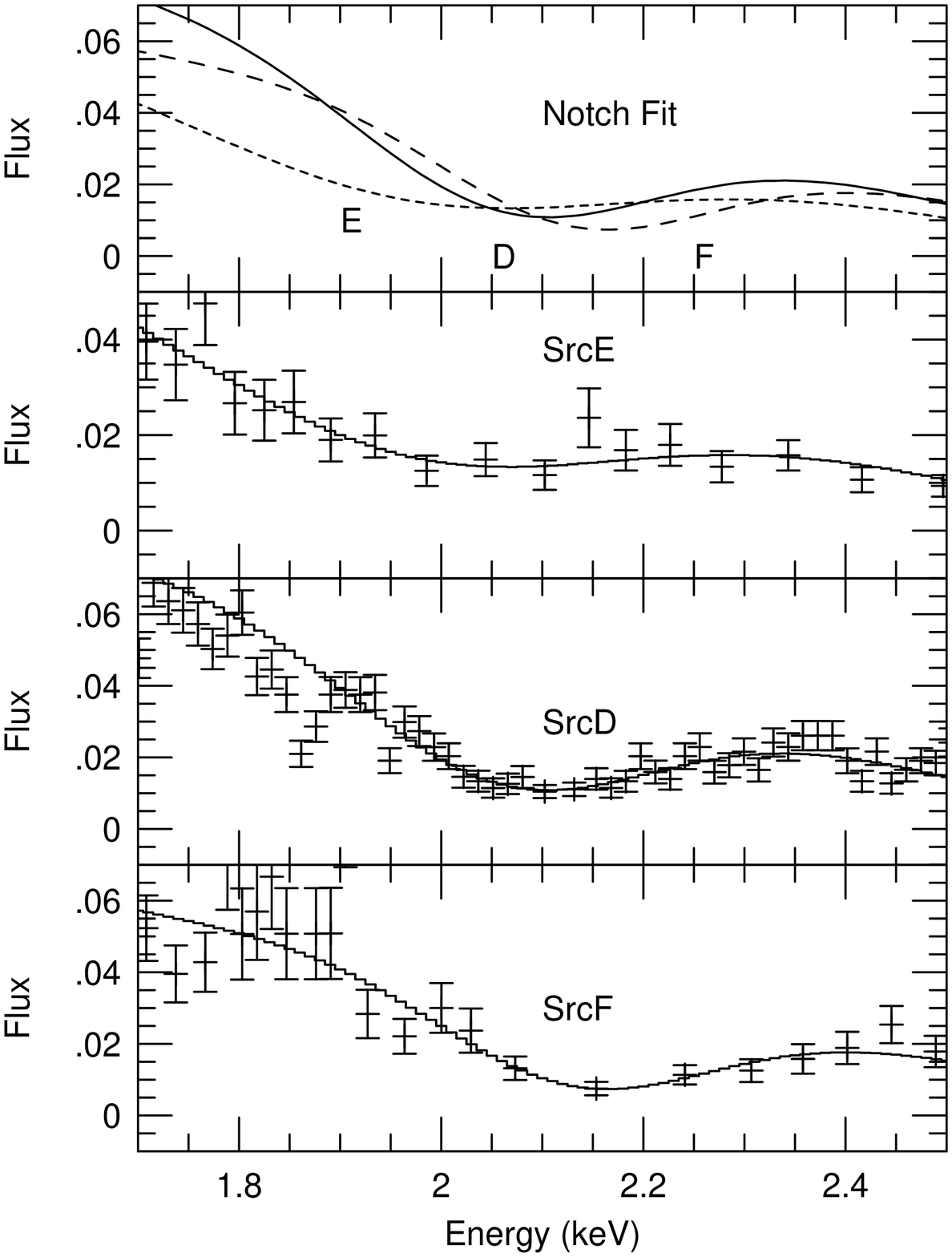]{The spectra around the instrumental 
absorption feature near 2.1 keV.  The 3 lower panels show the 
data and fit for sources D, E, and F.  The top panel compares the 
fits for the 3 sources.  
   \label{fig2}}
   
\figcaption[ecar_notch.ps]{The data from the fits to the 2.1 keV 
feature for sources D, E, and F.  The top panel shows the energy
of the line center as a function of distance from the chip readout.
The lower panel shows the line width as a function of distance from
the chip readout.
   \label{fig3}} 

\figcaption[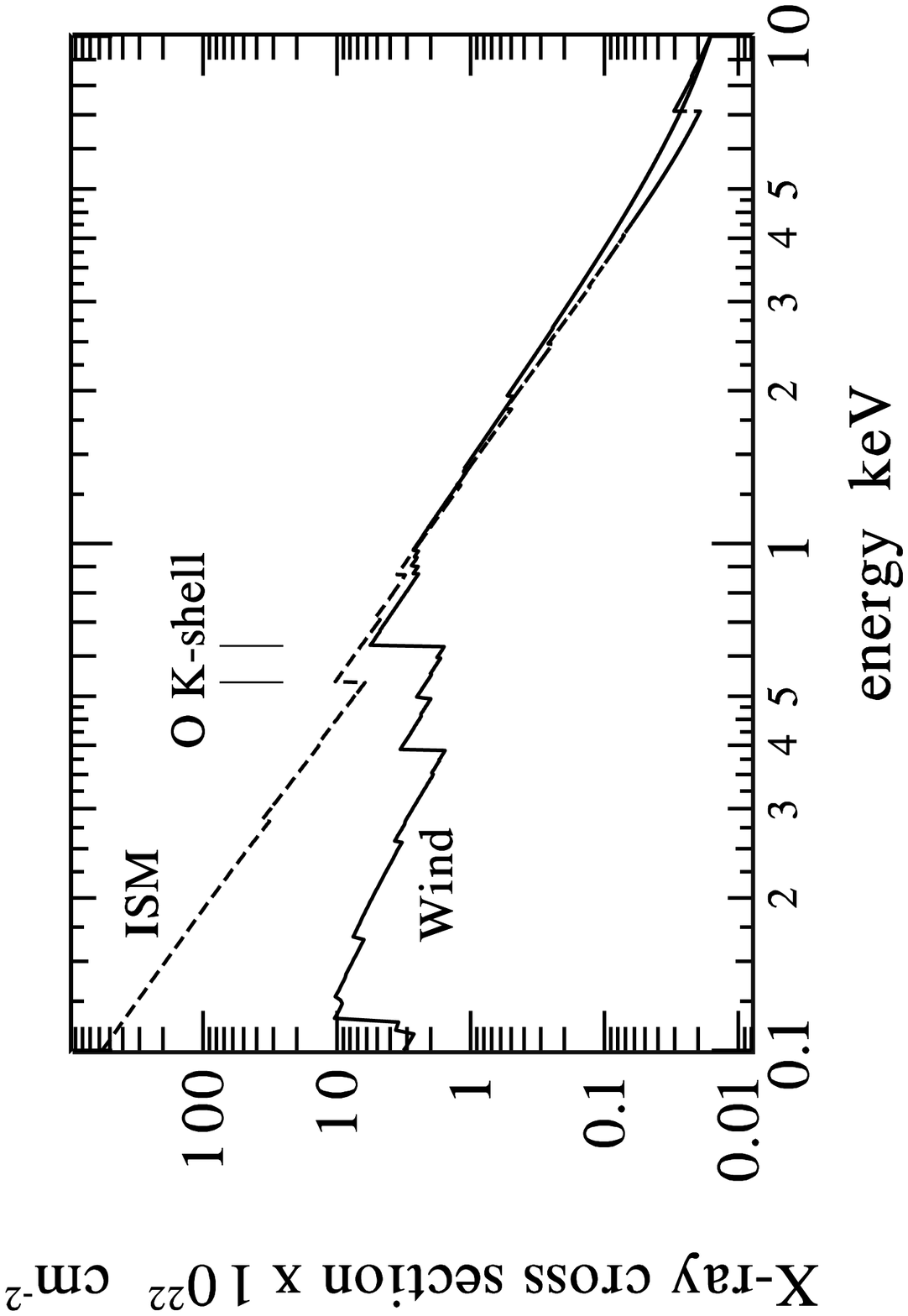]{Comparison of the energy 
dependent wind absorption cross sections with the 
ISM cross sections.  The wind cross sections were calculated according to  
Waldron et al. (1998) and the ISM cross sections are from Shull and Van      
Steenberg (1985).
   \label{fig102}}

\figcaption[93129fit.eps]{The spectrum of HD 93129AB (Source F).  Top: 
The data compared with the model.  Bottom: The three components from 
which the model in the top panel  is composed.  \label{fig101}}

\figcaption[93162fit.eps]{The spectrum of HD 93162 (Source D) compared 
with the model (top).  The three components from which 
the model is composed (bottom).    \label{fig103}}                                                            

\figcaption[93250fit.eps]{The spectrum of HD 93250 (Source E) compared 
with the model (top).  The two components from which 
the model is composed (bottom).    \label{fig103}}

\figcaption[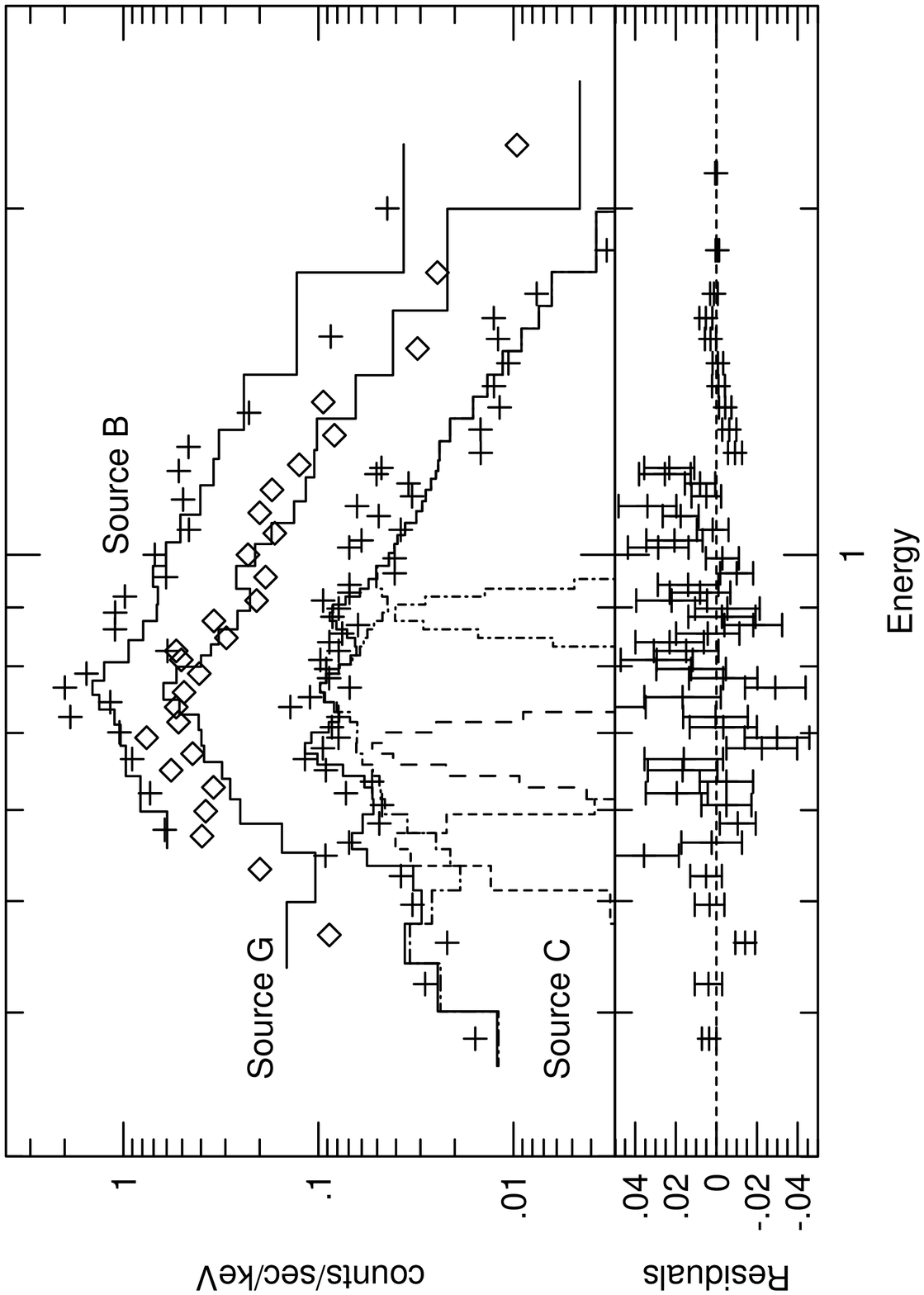]{(top) Combined plot for sources 
B (O6) +'s, C (O3) +'s, and G (O3) diamonds.
The errorbars have been suppressed for legibility of the spectra but
their size may be judged from the residuals plot (bottom).  Source B's
spectrum has been offset vertically by 1.6; source G's spectrum has
been offset vertically by 1.0.
The solid lines are the total model fit to
each spectrum.  For source C, that total model is composed of a
continuum plus 3 gaussian emission lines with each component shown as
a broken line.  The residuals are those of source C; the residuals for
sources B and G are similar in their vertical range.
   \label{fig101}}

\figcaption[spececg.ps]{The comparison of the spectra of O3 stars.
Spectrum E, the reference spectrum, is the solid line; source C is the 
squares; source G is the triangles. Spectra have been
scaled so that sources C and G have the same flux as source E 
near 1 keV.  
   \label{figa}}

\figcaption[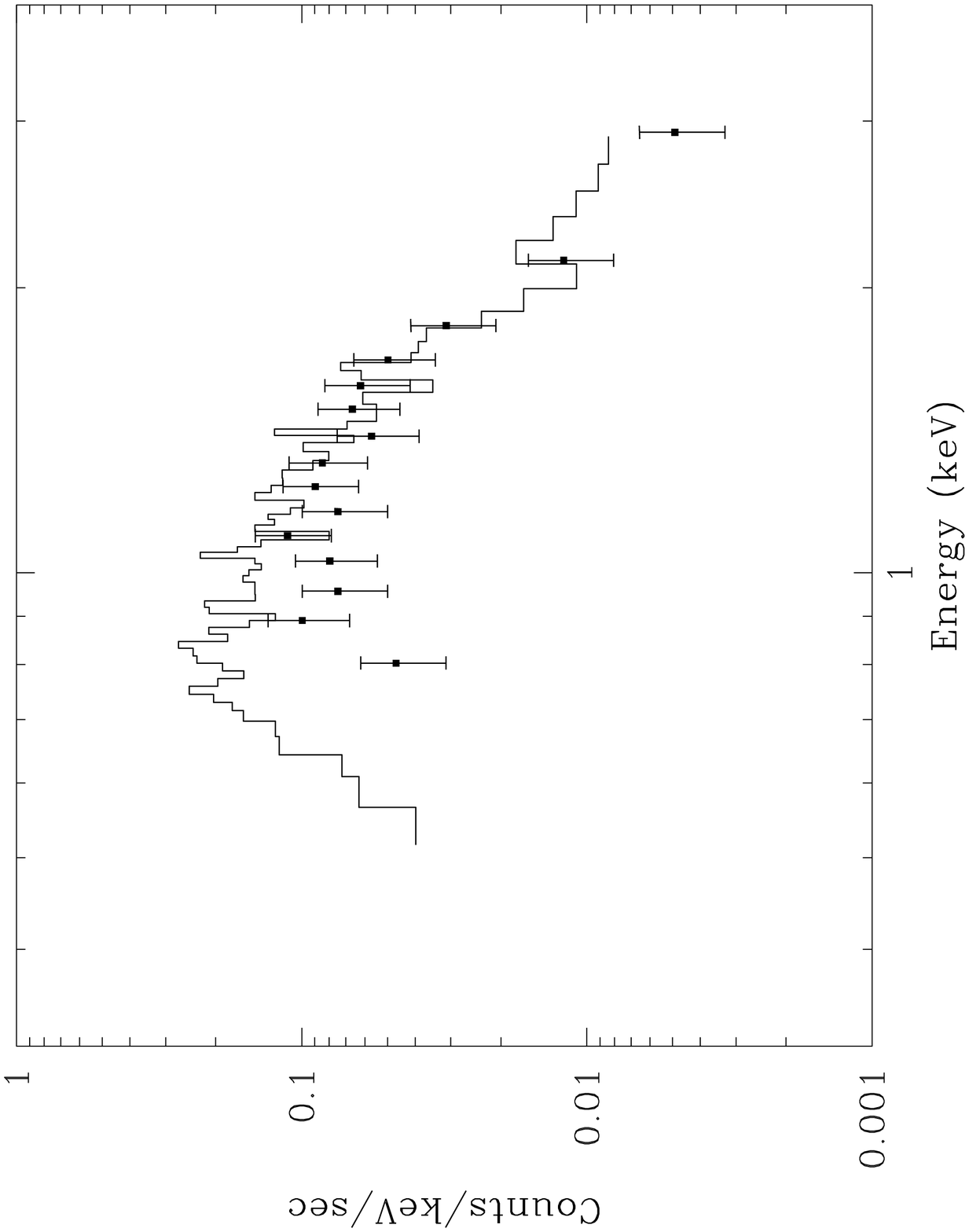]{The comparison of  O3 stars with different
optical absorption.
Spectrum E, the reference spectrum, is the solid line; source L  
(squares) is more heavily reddened.
   \label{figb}}

\figcaption[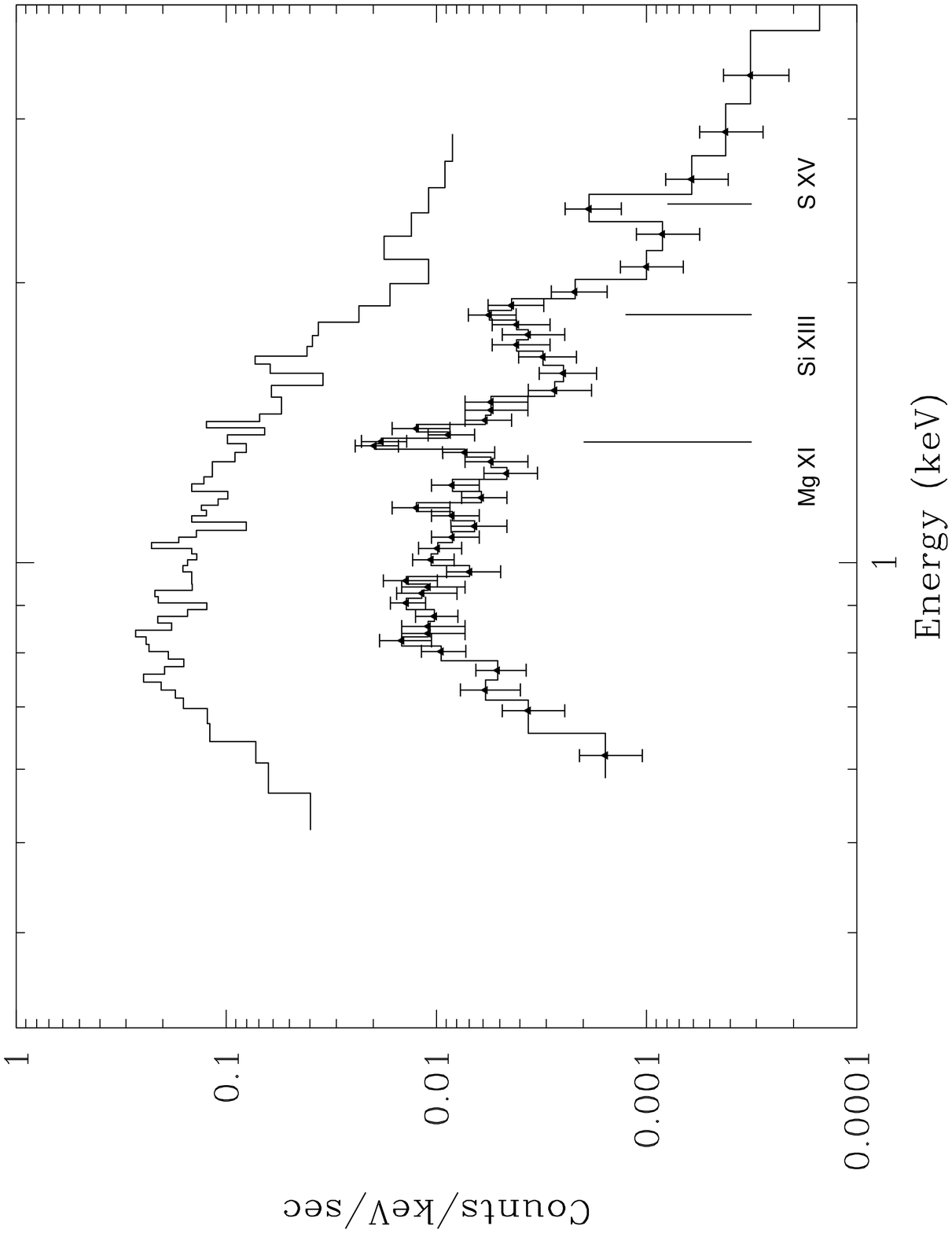]{The O3 f star.
Spectrum E, the reference spectrum, is upper spectrum; source F is the lower
spectrum. The 2 spectra have been offset to display the relatively smooth
spectrum of E and the lines in source F. 
\label{figc}}

\figcaption[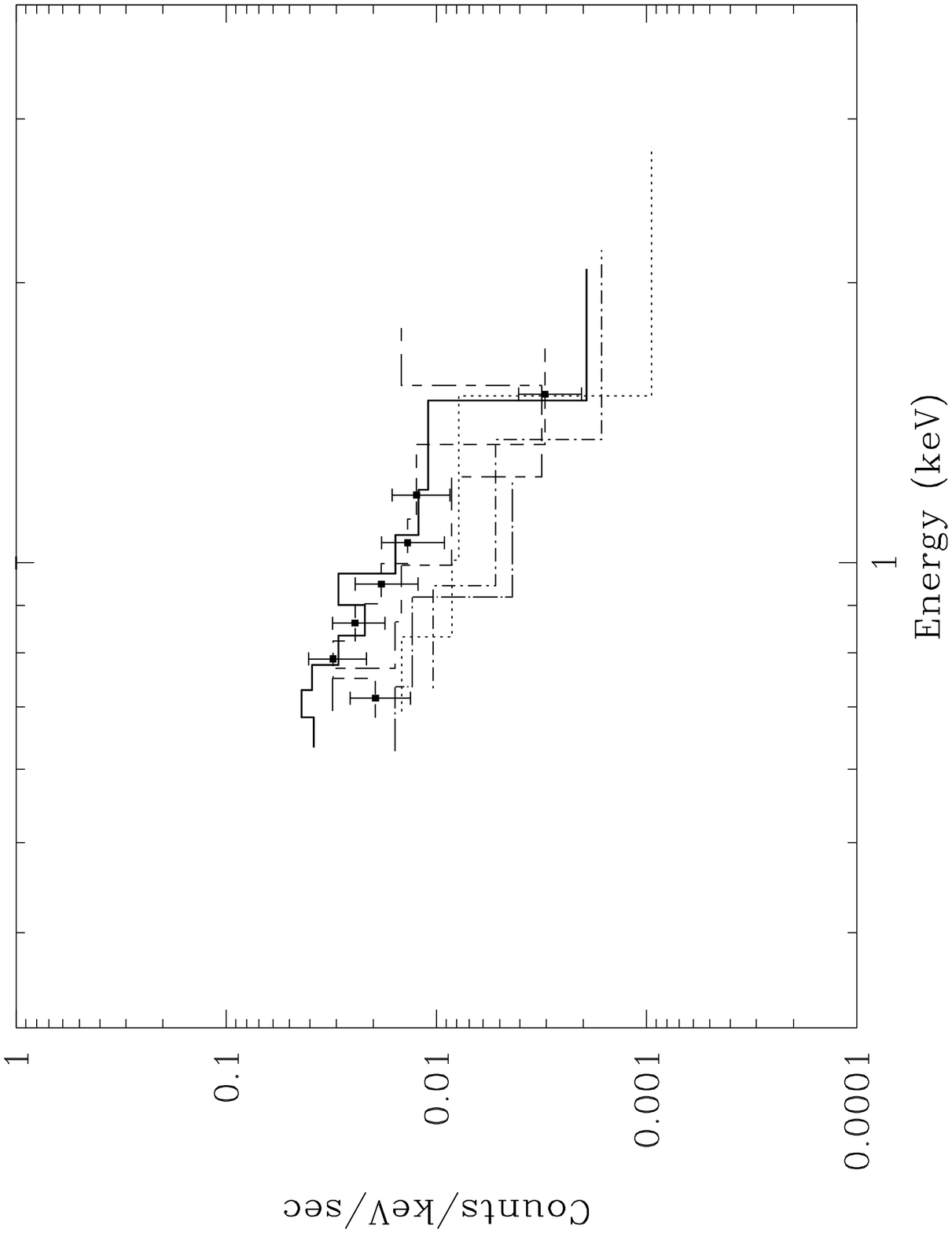]{The comparison of the spectra of the 
late O stars.  Source B: heavy solid line; source H: short dash line;
source M: dotted line; source N dot short dash line; source O: dot long dash
line; source P: short dash--long dash line. Error bars are 
shown for source H;  error bars for other sources are comparable.
Sources have $\it not$ been scaled.  
\label{figd}}

\figcaption[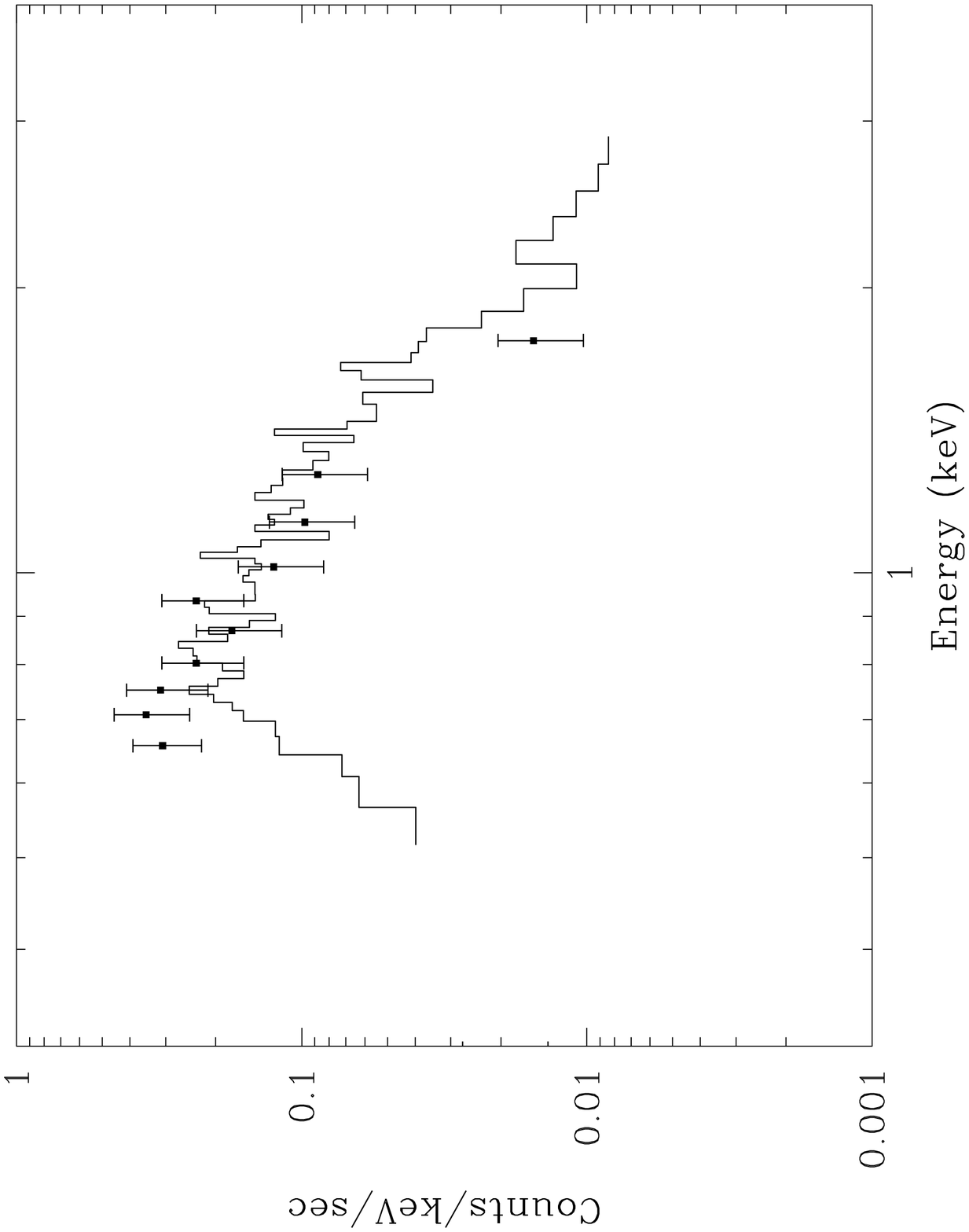]{The comparison of the spectra of  O3 stars
with an O6 star (source B).  a.  The O3 standard spectrum
(source E, histogram) and an O6 star (source B, squares).
b. Source C (O3 V, histogram) and an O6 star (source B, squares). 
\label{fige}}

\figcaption[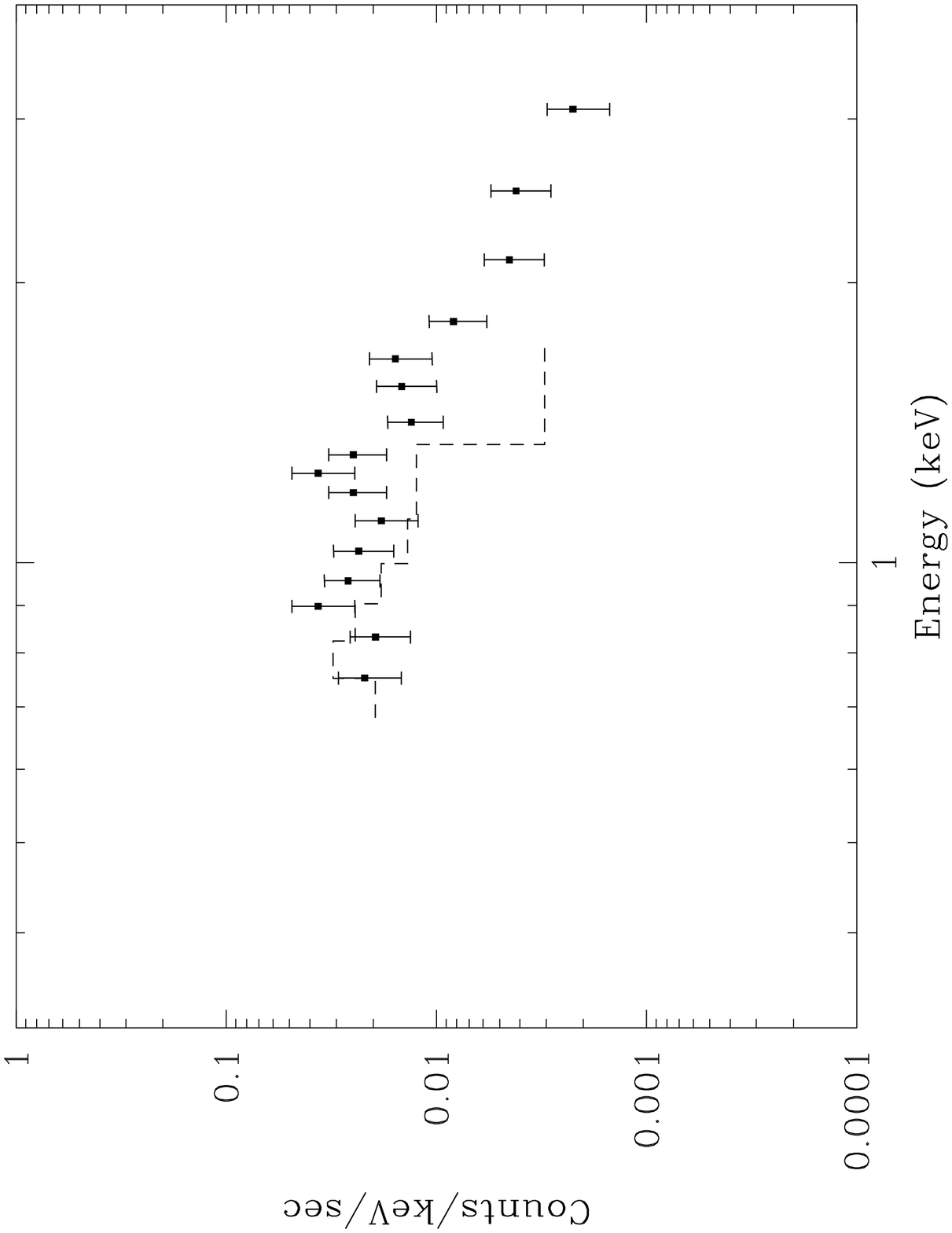]{The comparison of the spectra of an O8.5  star
with extra reddening (source A, squares with error bars) and an O8 star with a 
reddening typical of other cluster stars (source H, dashed line).
\label{figf}}

\figcaption[specega.ps]{The comparions of the Wolf-Rayet
star (Source D, showing error bars) with Source E (dashed line), 
the O3 standard star. a. The counts have not been scaled.  The 
dip near 2. keV in both spectra is an instrumental artifact. 
Apart from that dip, both spectra lack strong features.
b. In this case, the counts for the Wolf-Rayet star have been                          
scaled (by a factor of 8).
\label{figga}}


\figcaption[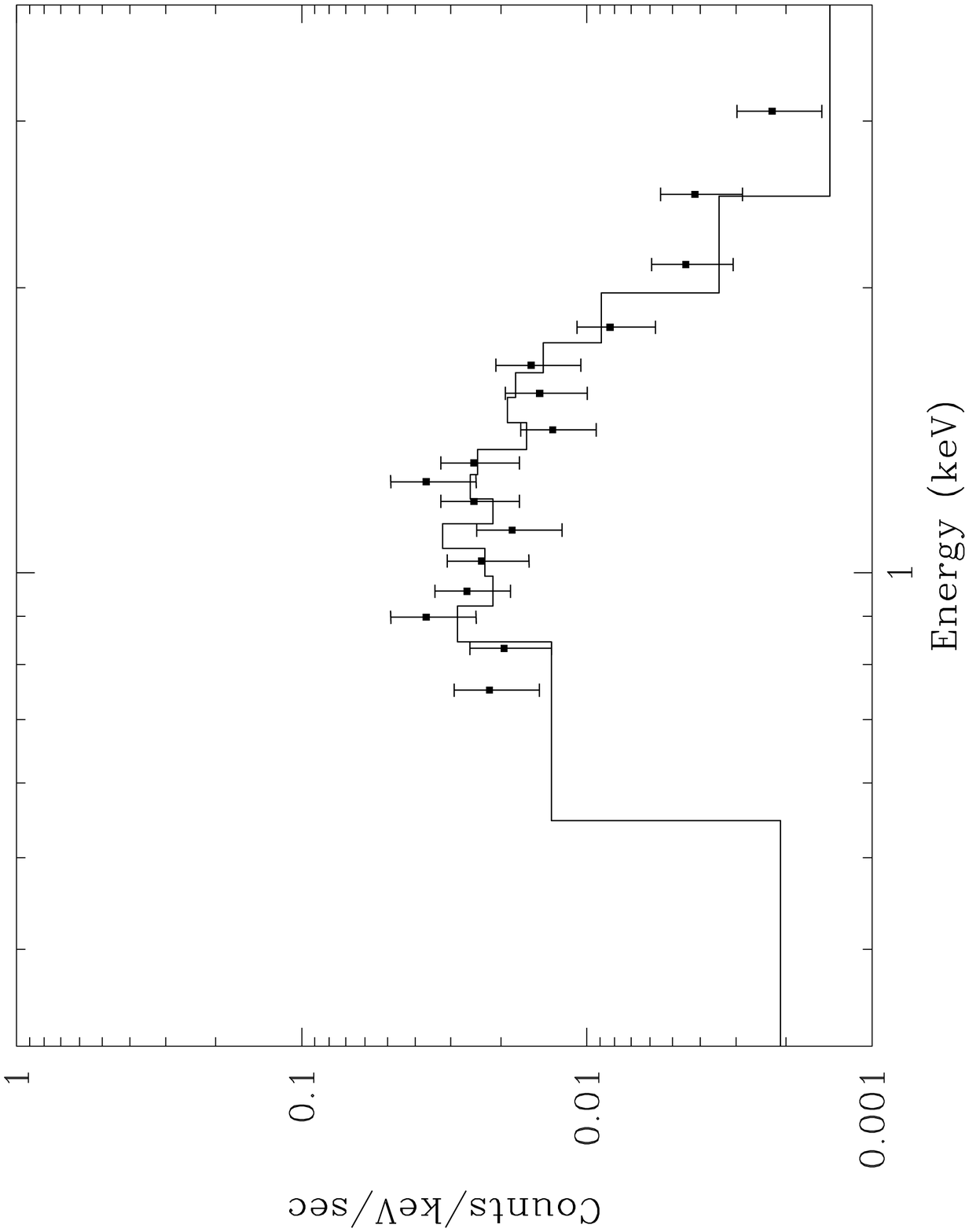]{Two heavily reddened sources.  Source L: solid
histogram; Source A squares.  Despite very different photospheric 
spectral types (source L: O3 I; source A: O8.5 V), the X-ray spectra
are identical. Source A is also shown in Fig. 14, which illustrates
that it is an absorbed spectrum. 
\label{figh}} 

\figcaption[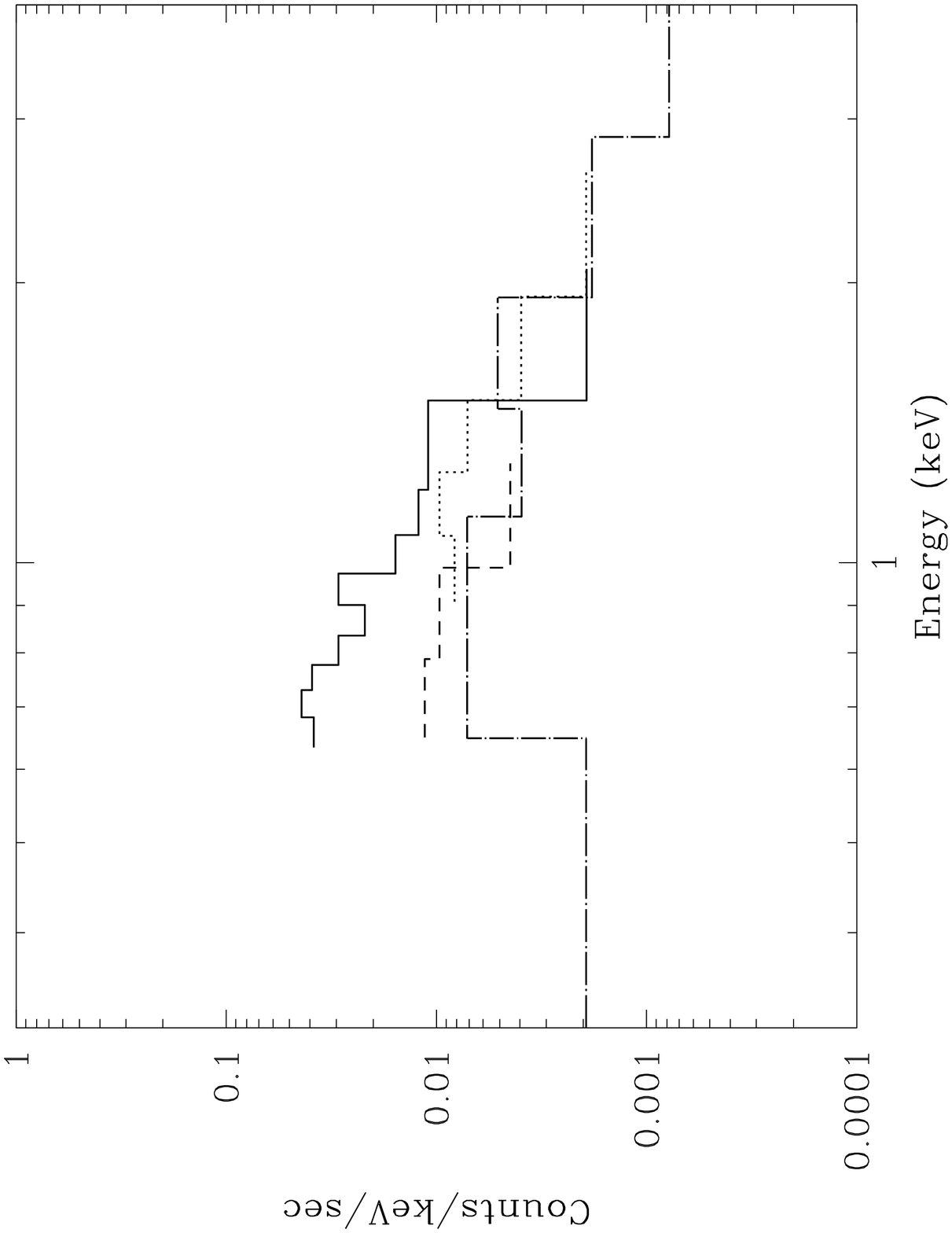]{Spectra of 3 cool stars (broken lines:
source I: dot-dash line; source J dashed line; source K: dotted line).
For comparison, the spectrum of  a representative hot star is shown
(source B: solid line).  The cool stars have flatter, harder spectra than
the hot star. 
\label{figi}} 

\figcaption[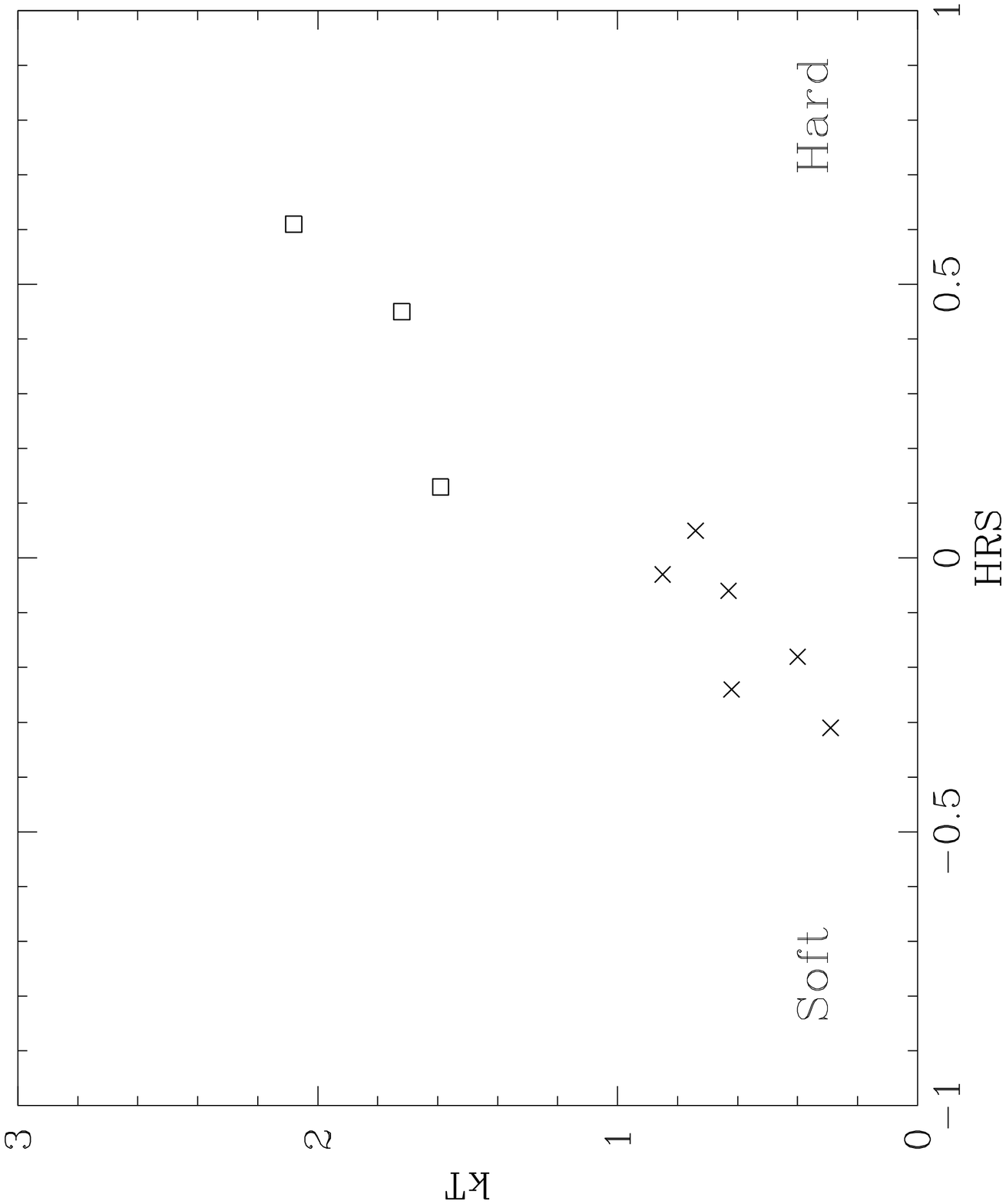]{Temperatures and hardness ratios for the
stronger sources (squares) and the weaker sources (x's).
 The temperatures for the ``restricted"
fits to the weaker hot stars for  N$_H$ = 3 x 10$^{21}$ cm$^{-2}$. 
HRS is the hardness ratio for the energy bands 0.5 to 0.9 keV 
and 0.9 to 1.5 keV (HR$_{MS}$); kT is the temperature in keV.  For the stronger
sources (D, E, F), temperatures are the mean of the two 
temperatures from Table 5.  
\label{fig }} 

\figcaption[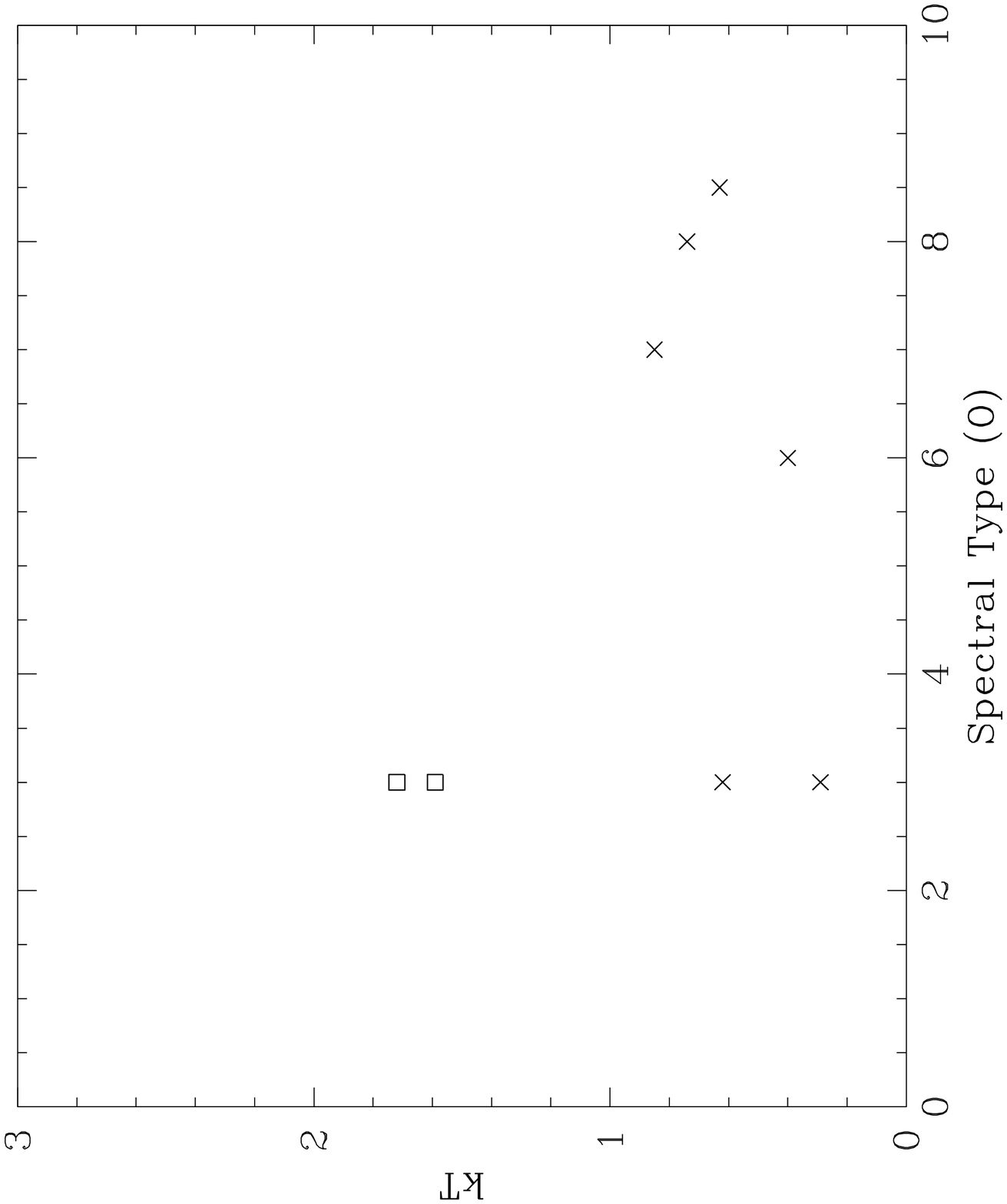]{The temperatures of the ``restricted"
fits to the weaker hot stars for  N$_H$ = 3 x 10$^{21}$ cm$^{-2}$. 
Spectral type is the spectral subclass for O stars; 
kT is the temperature in keV. For the stronger sources (E and F, squares),
temperatures are the mean of the two temperatures 
 from Table 5.  The Wolf-Rayet star (Source D) is 
omitted because its spectral type is not comparable.)  
\label{fig }} 

\figcaption[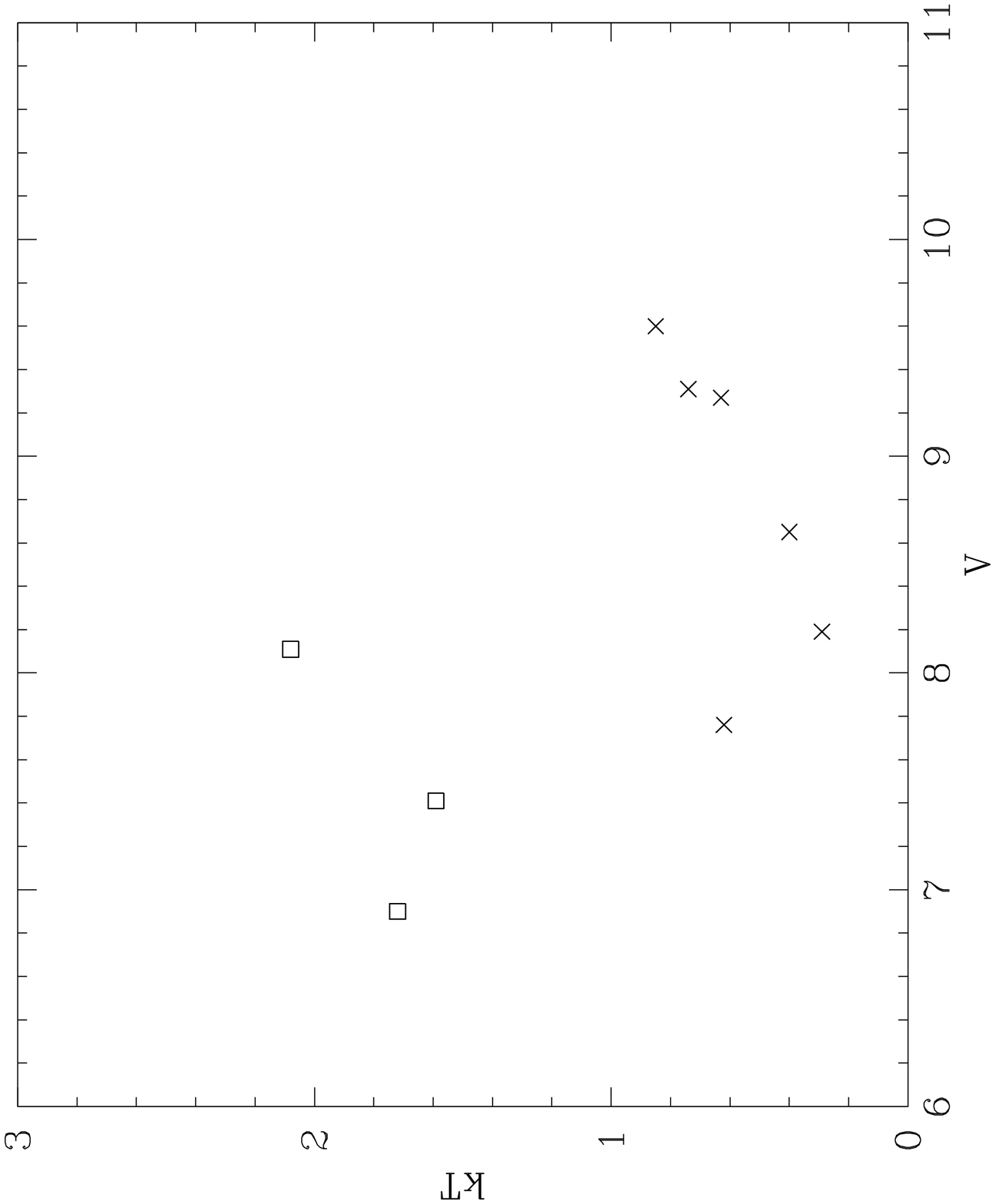]{The temperatures of the ``restricted"
fits to the weaker hot stars for  N$_H$ = 3 x 10$^{21}$ cm$^{-2}$. 
V is the magnitude (uncorrected for absorption):
kT is the temperature in keV.  For the stronger sources (D, E, F, 
squares), temperatures the mean of the two temperatures  from Table 5.
\label{fig }} 

\plotone{srcspec.ps}

\plotone{ecar_notch.ps}

\plotone{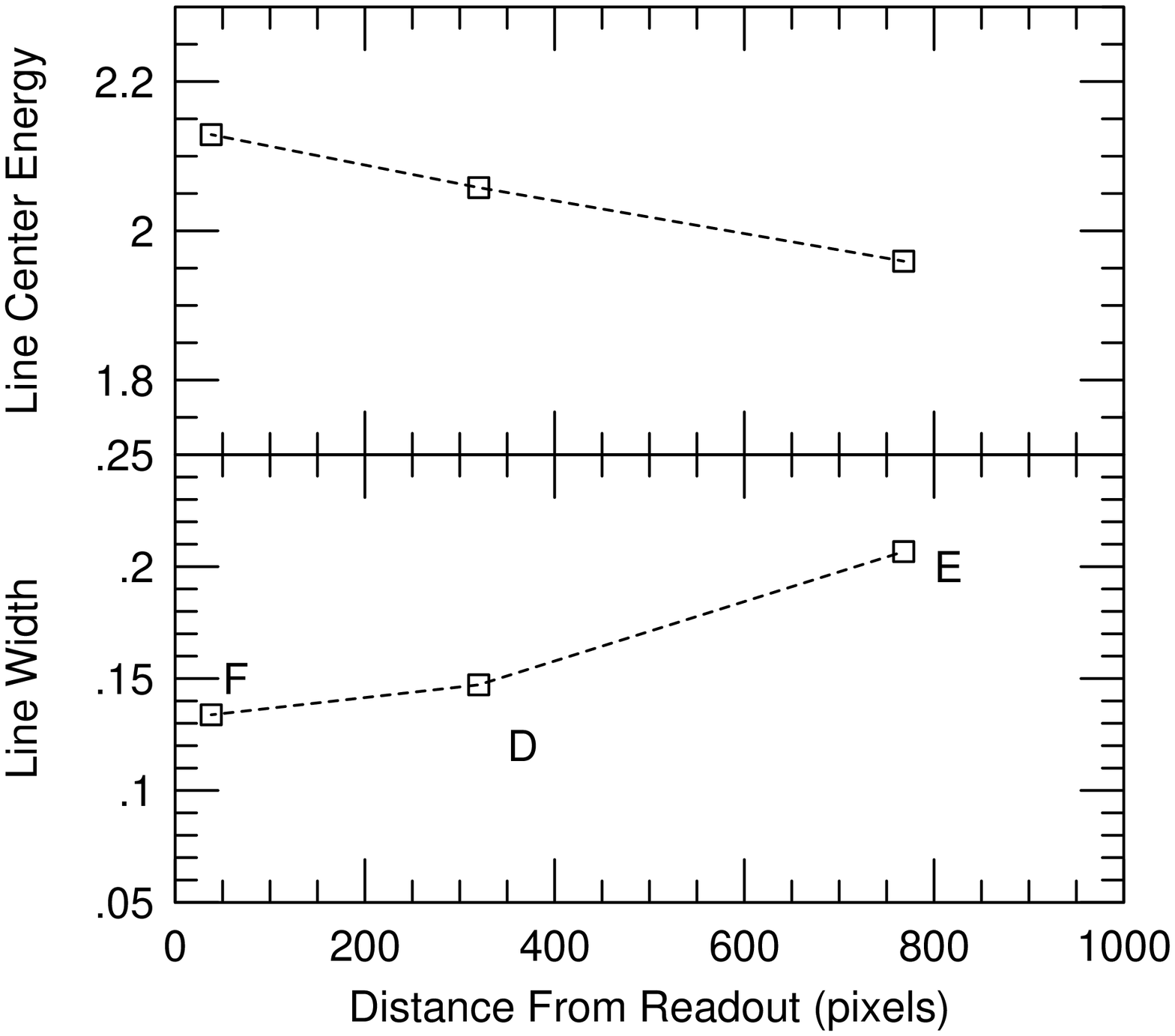}

\plotone{Opismw.eps}



\plotone{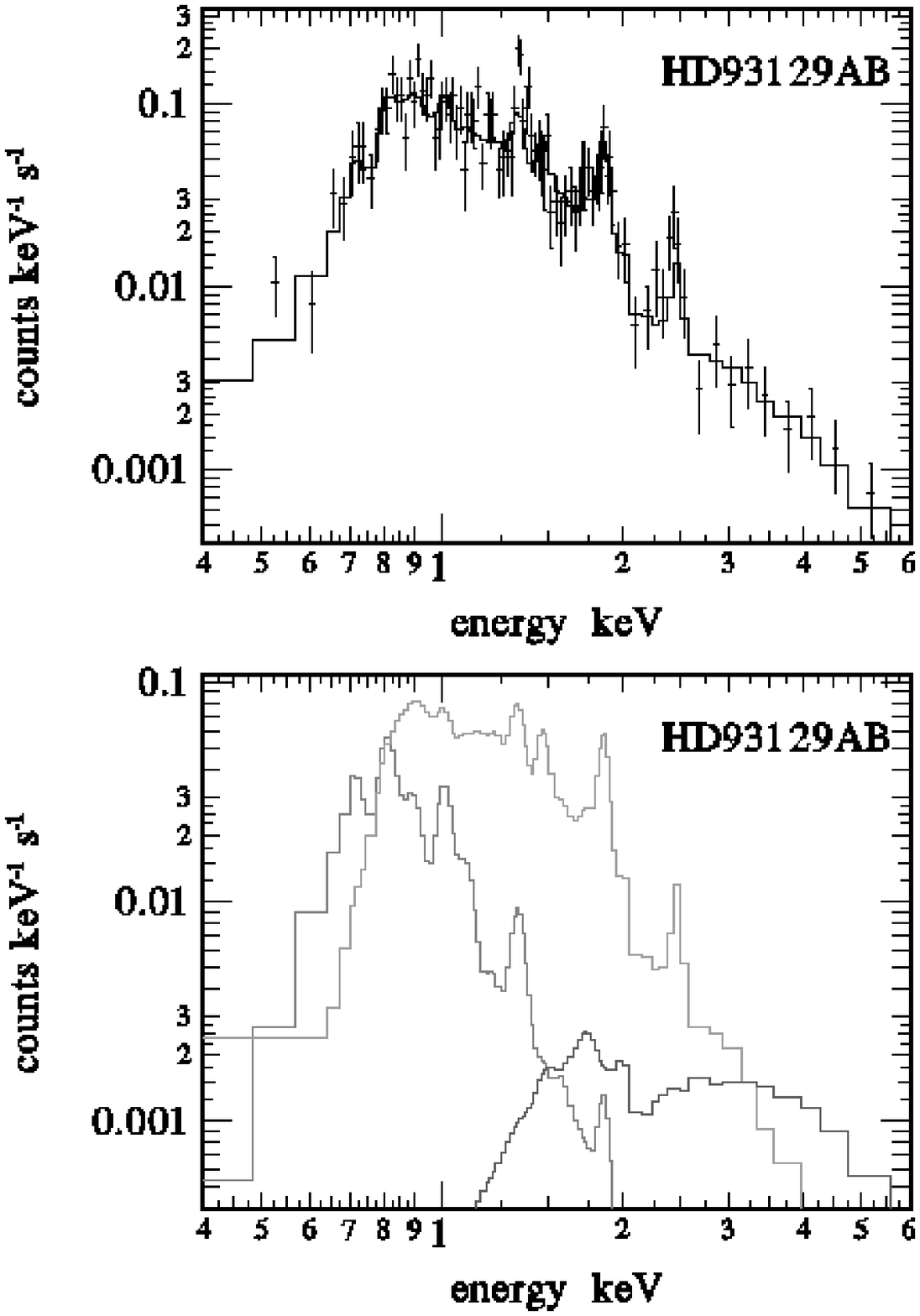}

\plotone{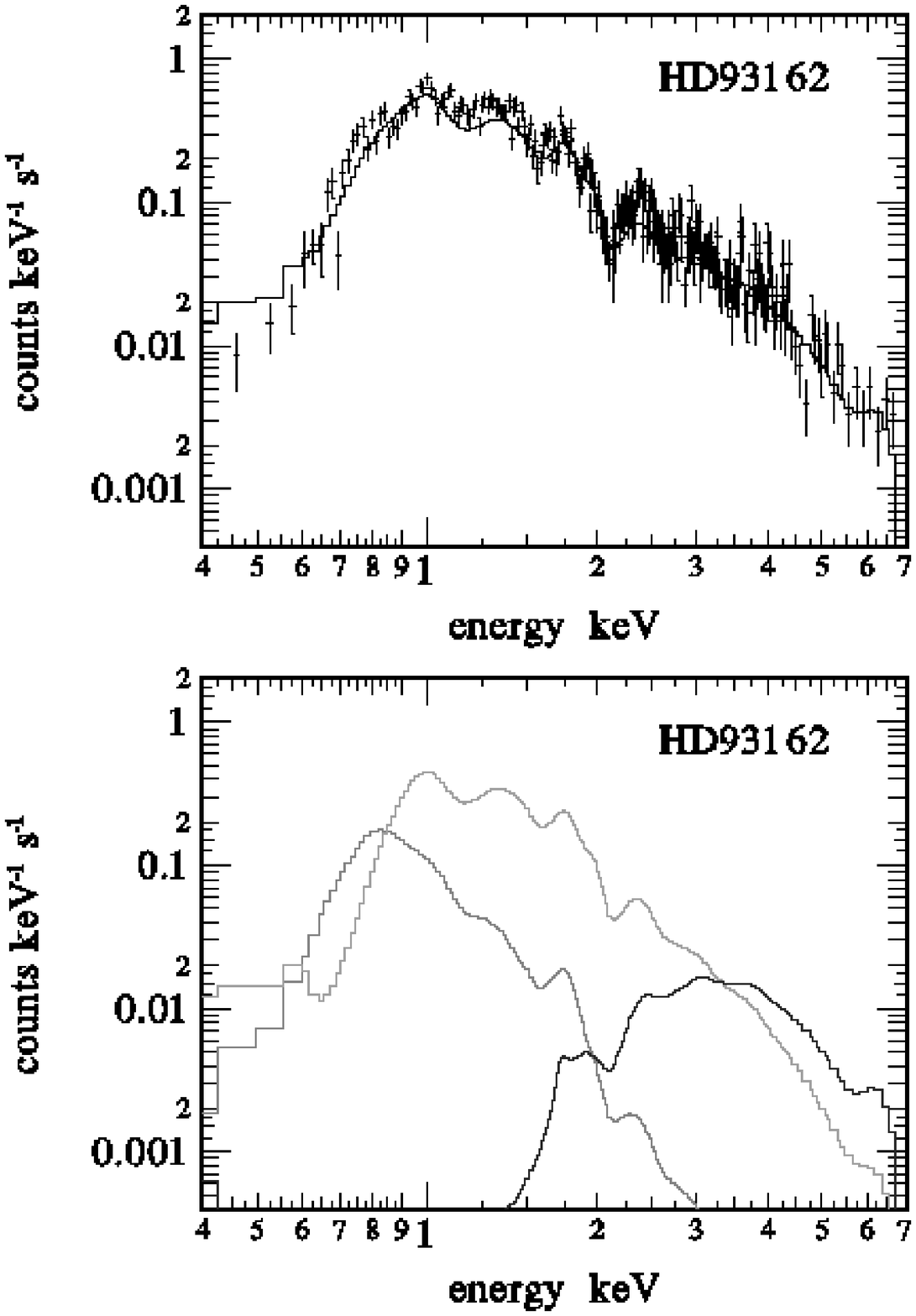}

\plotone{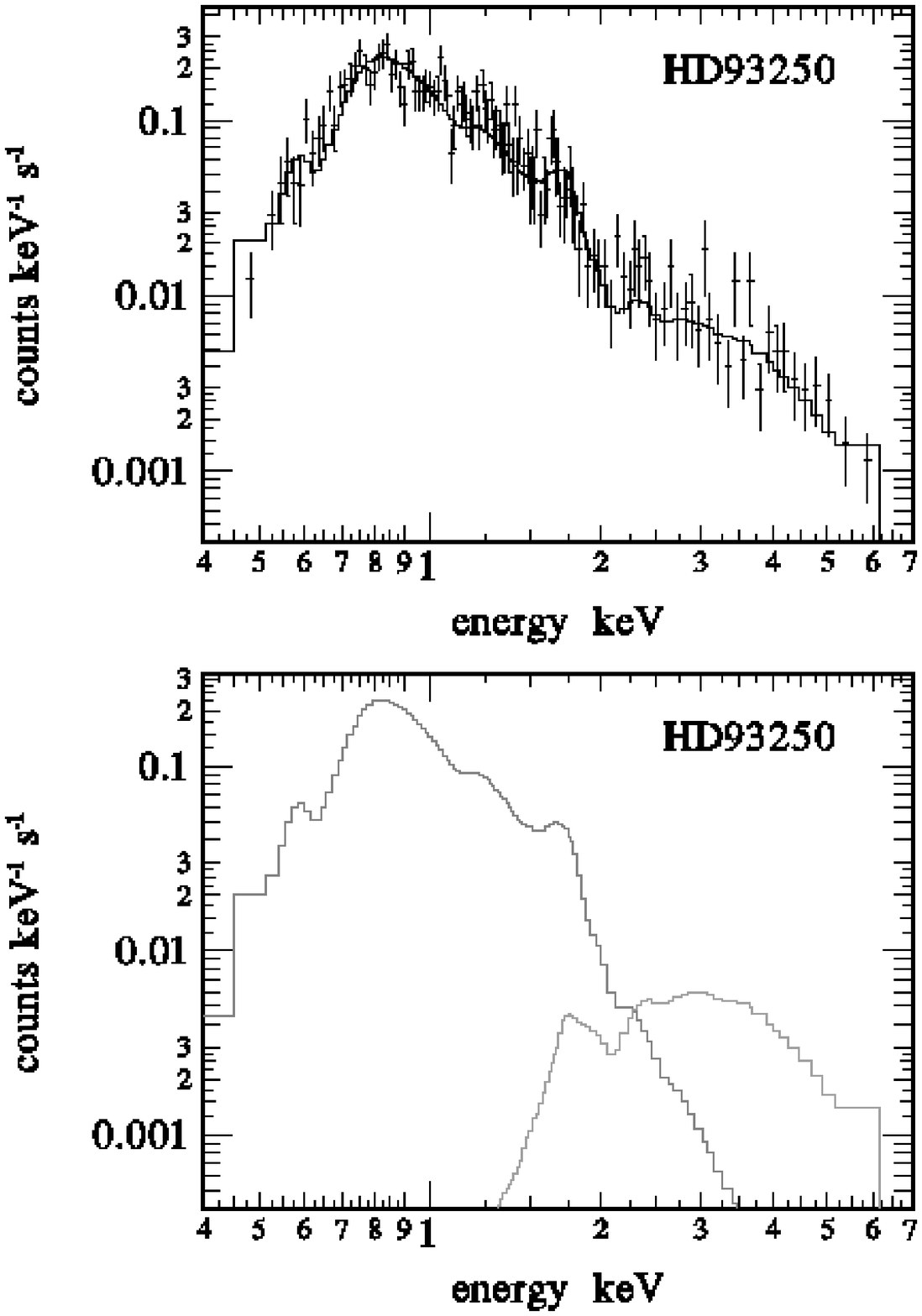}





\plotone{comb_spec_v2.ps}


\plotone{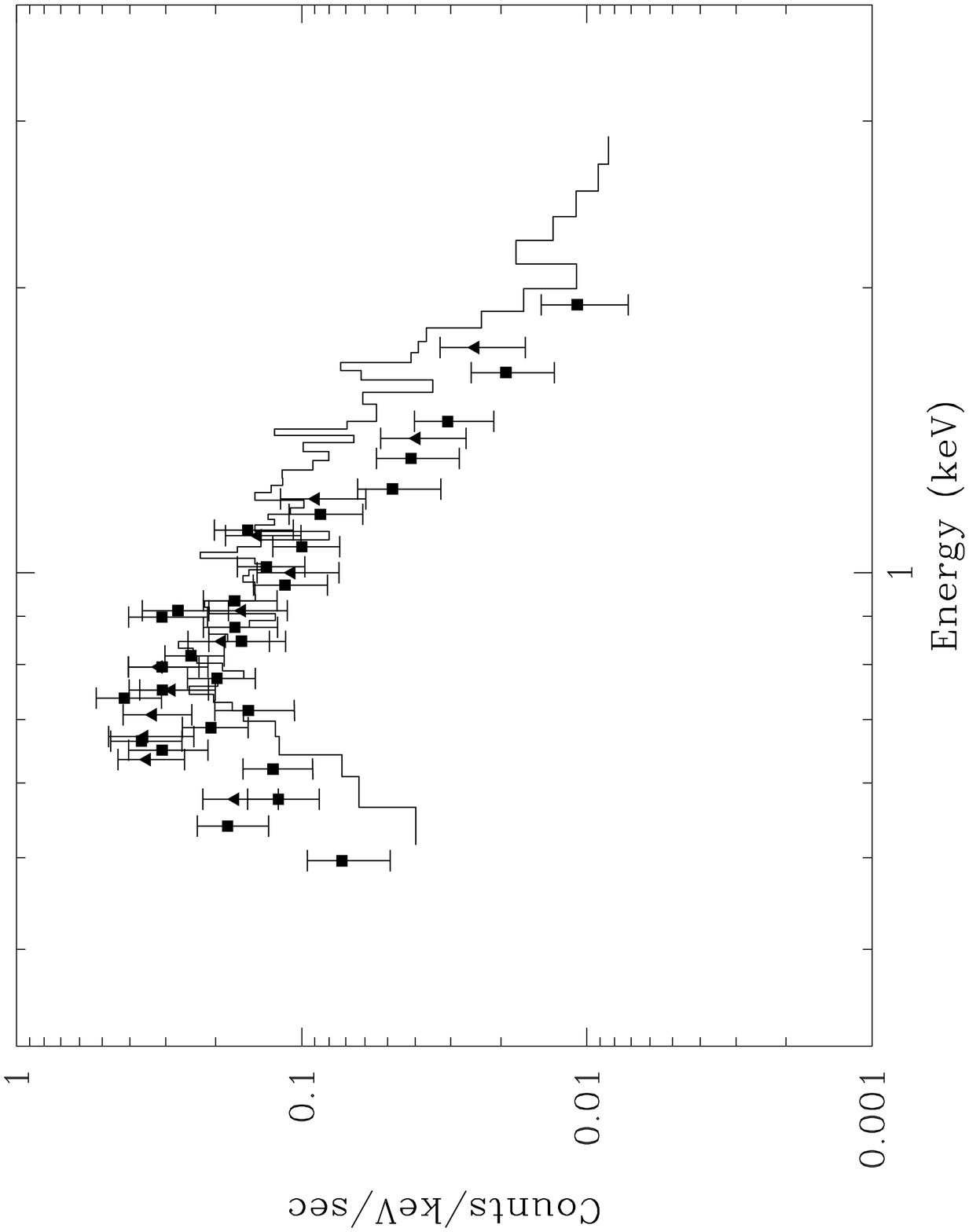}


\plotone{spececb.ps}

\plotone{spececc.ps}

\plotone{spececd.ps}


\plottwo{specece.ps}{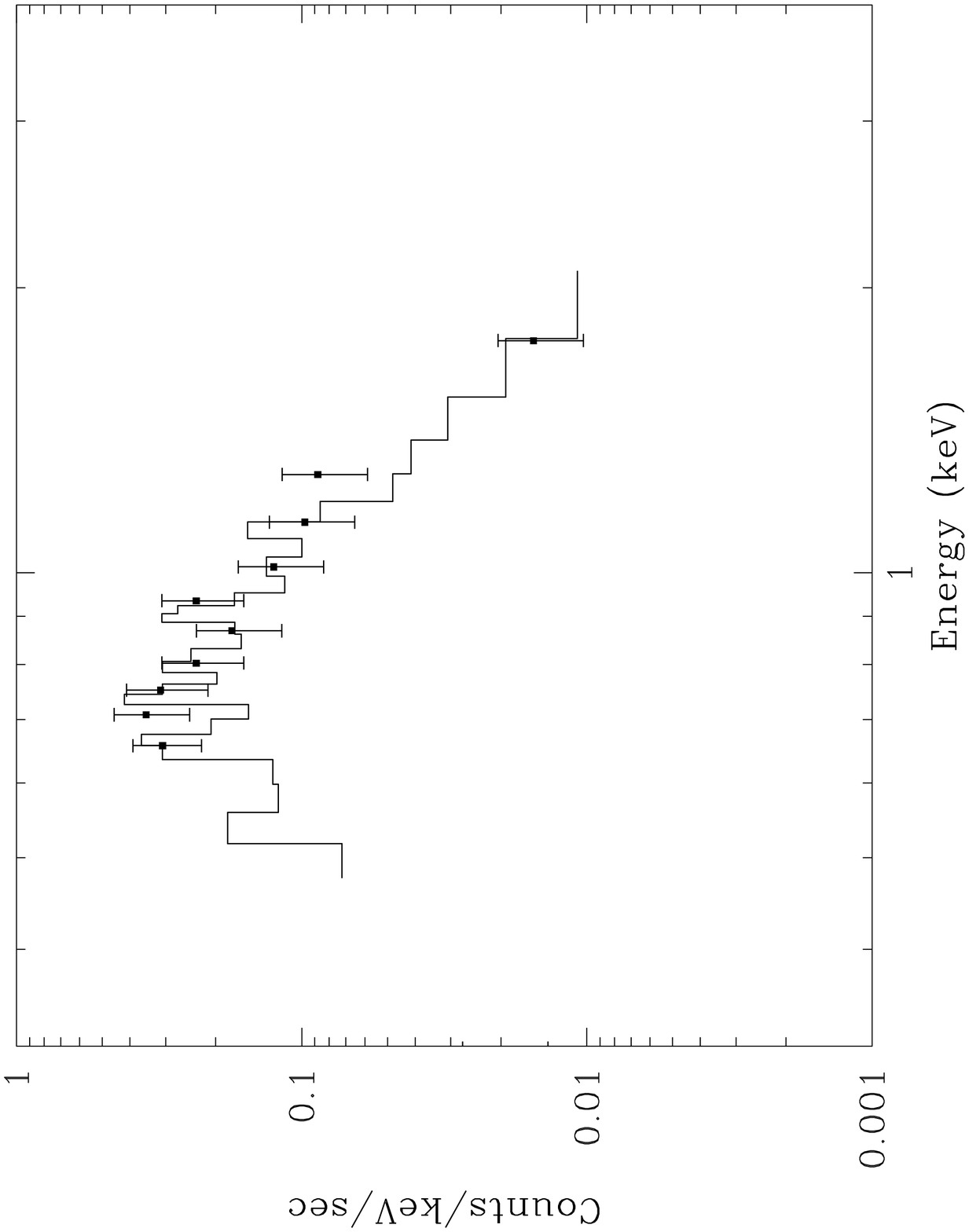}


                                                                                
\plotone{spececf.ps} 


\plottwo{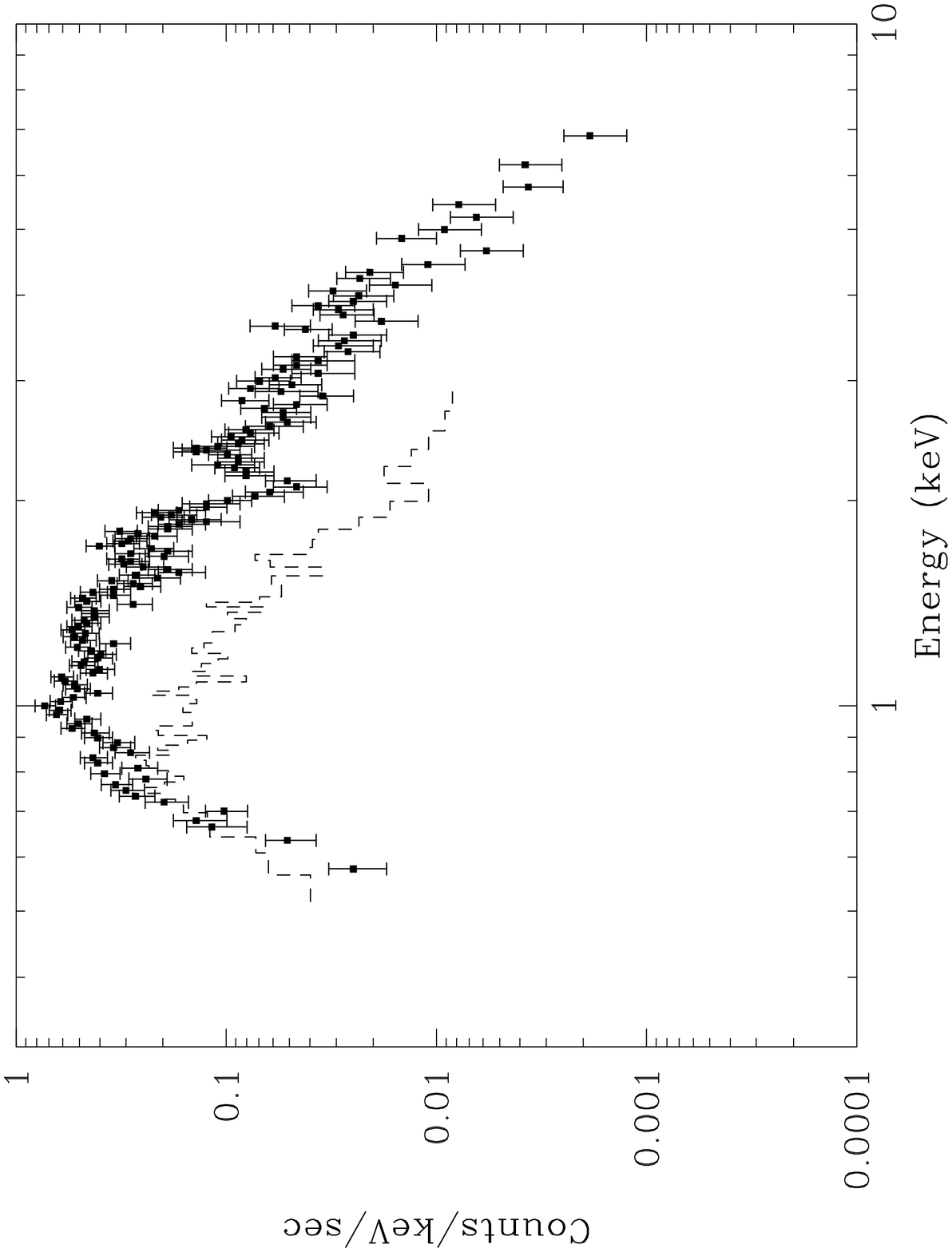}{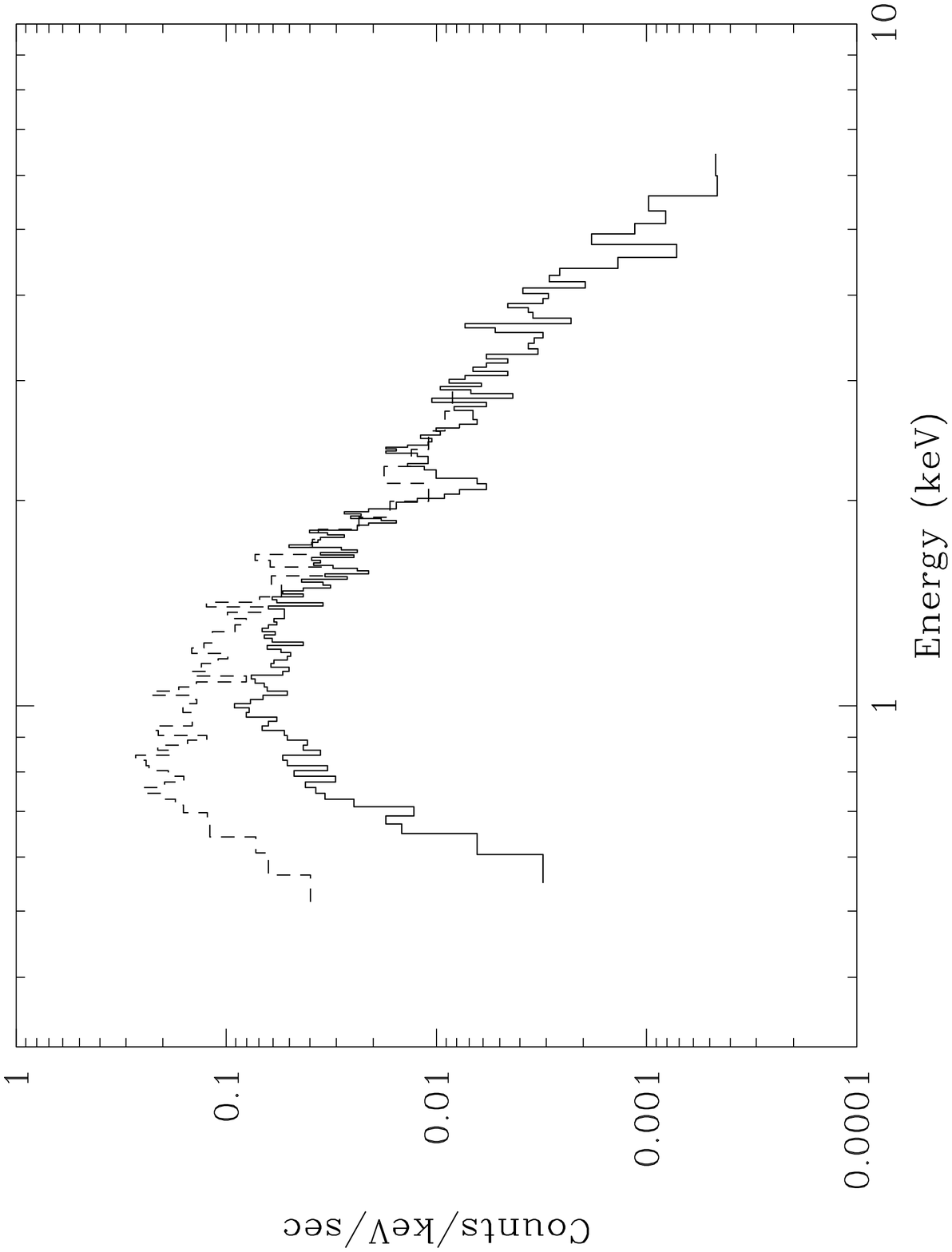}
                                                                                


\plotone{specal.ps} 

\plotone{spececi.ps}

\plotone{thra.ps}

\plotone{tspa.ps}

\plotone{tva.ps}



\clearpage



\begin{thebibliography}{}


\bibitem[Albacete etal]{al02}Albacete Colombo, J. F., Morrell, N. I., 
Rauw, G., Corcoran, M. F., Niemla, V. S., and Sana, H. 2002, \mnras, 
336, 1099


\bibitem[Albacete etal]{al03}Albacete Colombo, J. F., Mendez, M., and 
Morrell, N. I. 2003, \mnras, in press

\bibitem[Benaglia and Koribalski]{bk04}Benaglia, P. and Koribalski, B 
2004, \aap. in press

\bibitem[Cassinelli and Olson]{co79}Cassinelli, J. P., and Olson, G. L. 1979, ApJ, 229, 304  

\bibitem[Cassinelli, et al 2001]{ca01}Cassinelli, J., P., 
Miller, N. A., Waldron, W. L., MacFarlane, J. J., and Cohen, D. H.
2001, \apj, 554, L55

\bibitem[Corcoran, et al 2001]{co01}Corcoran, M. F., Swank, J. H., Petre, R.,
Ishibashi, K., Davidson, K., Townsley, L., Smith, R., White, S., Viotti, R.,
and Damineli, A. 2001, \apj, 562, 1031



\bibitem[Evans, et al, 2003]{ev03}Evans, N. R. Seward, F. D., Krauss, M. I.,
Isobe, T., Nichols, J. Schlegel, E. M., and Wolk, S. J. 2003, \apj, 
589, 509 (Paper I)


\bibitem[Garmire, et al, 2003]{ga03}Garmire, G. P., Bautz, M. W.,
Ford, P. G., Nousek, J. A., and Ricker, G. R. 2003, SPIE, 4851, 28






\bibitem[Howarth and Prinja, 1987]{hp89}Howarth, I. D., and 
Prinja, R. K. 1989, \apjs, 69, 527 

\bibitem[Kahn, et al 2001]{ka01}Kahn, S. M., Leutenegger, M. A., Cottam,
J., Rauw, G., Vreux, J.-M., Boggende, A. J. F. den, Mewe, F., and G\"udel, 
M. 2001, \aap, 365, L312

\bibitem[Lamers and Leitherer 1993]{ll93}
Lamers, H. J. G. L. M., and Leitherer, C. 1993, \apj, 412, 771

\bibitem[MacFarlane etal 1991]{Mf91}
MacFarlane, J. J., Cassinelli, J. P., Welsh, B. Y., Vedder, P. W., Vallerga,    
J. V., and Waldron, W.                                                            
L. 1991, \apj, 380, 564                                                          
  


\bibitem[Massey and Johnson 1993]{mj93}Massey, P., and Johnson, J. 1993, 
\aj, 105, 980   


\bibitem[Miller etal 2002]{mi02}
Miller, N. A., Cassinelli, J. P., Waldron, W. L., MacFarlane, J. J., and          
Cohen, D. H. 2002, \apj, 577, 951                                                                        
 


\bibitem[Raassen, etal]{ra03}Raassen, A. J. J., van der Hucht, K. A., Mewe,
R., Antokhin, I. I., Rauw, G., Vreux, J.-M.,, Schmutz, W., and G\"udel, M.
2003, \aap, 402, 653


\bibitem[Schulz, et al 2003]{sc03}Schulz, N. S., Canizares, C., 
Huenemoerder, D., and Tibbets, K. 2003, \apj, in press


\bibitem[Seward, 2000]{fds00}Seward, F. D. 2000 in {\it Astrophysical 
Quantities}, ed. A. N. Cox (New York: Springer Verlag), p 197

\bibitem[Seward, et al 2001]{se01}Seward, F. D., Butt, Y. M., Karovska, M., Prestwich, A.,
and Schlegel, E. M. 2001, \apj, 553, 832

\bibitem[Shull and van Steenberg,1985]{svs85}
Shull, J. M., and Van Steenberg, M. E. 1985, \apj, 294, 599

\bibitem[Skinner, et al 2001]{Sk01}Skinner, S. L., G\"udel, M.,
Schmutz, W., and Stevens, I. R. 2001, \apj, 558, L113

\bibitem[Stevens, et al 1996]{St96}Stevens, I. R., Corcoran, M. F., 
Willis, A. J., Skinner, S. L., Pollock, A. M. T., Nagase, F., 
and Koyama, K. 1996 \mnras, 283, 589  


\bibitem[Townsley, et al 2000]{To00}
Townsley, L. K., Broos, P. S., Garmire, G. P., and Nousek, J. A.
2000, \apj, 534, L139

\bibitem[Waldron 1984]{wa84}Waldron, W. L. 1984, ApJ, 282, 256

\bibitem[Waldron and Cassinelli 2001]{wc01}Waldron, W. L., and Cassinelli, J. P. 
2001, ApJ, 548, L45

\bibitem[Waldron etal 1998]{waetal98}Waldron, W. L., Corcoran, M. F., Drake, S. A., 
and Smale, A. P. 1998, ApJS, 118, 217

\bibitem[Walborn et al 2002]{wa02}Walborn, N. R., Howarth, I. D.,               
Lennon, D. J., Massey, P., Oey, M. S., Moffat, A. F. J.,                        
Skalkowski, G., Morrell, N. I., Drissen, L., and Parker, J. W.                  
2002, \aj, 123, 2754                                                            

\bibitem[Willis, et al 1995]{Wi95}Willis, A. J., Schild, H., Stevens, I. R.
1995, \aap, 298, 549 




\end{thebibliography}
\end{document}